\documentclass[12pt]{report}

\usepackage{graphicx, color}
\usepackage[utf8]{inputenc}
\usepackage[english]{babel}
\usepackage[a4paper,margin=2cm]{geometry}
\usepackage[colorlinks=true,linkcolor=red, citecolor=blue]{hyperref}
\usepackage{hyperref}
\usepackage{comment}
\usepackage{amsfonts}
\usepackage{amsmath}
\usepackage{bbm}
\usepackage{listings}
\usepackage{float}
\usepackage[table,xcdraw]{xcolor}
\usepackage[ruled, linesnumbered]{algorithm2e}
\usepackage{booktabs,array}
\usepackage{subfig}

\usepackage{ACL2023}

\begin{document}

\begin{titlepage}

\newcommand{\HRule}{\rule{\linewidth}{0.5mm}} 

\center 




\textsc{\Large MSc Artificial Intelligence}\\[0.2cm]

\textsc{\Large Master Thesis}\\[0.5cm]




\HRule \\[0.4cm]

{ \huge \bfseries Dr. Boot: Bootstrapping Program Synthesis Language Models to Perform Repairing}\\[0.4cm] 

\HRule \\[0.5cm]




by\\[0.2cm]

\textsc{\Large Noah van der Vleuten}\\[1cm]




{\Large July 26, 2023}\\[1cm] 

{48EC}\\ %

{November 2022 - July 2023}\\[1cm]%




\begin{minipage}[t]{0.4\textwidth}

\begin{flushleft} \large

\emph{Supervisors:} \\

{N. \textsc{Butt} MSc } \\ 
{M. \textsc{Macfarlane} MSc }

\end{flushleft}

\end{minipage}

~

\begin{minipage}[t]{0.4\textwidth}

\begin{flushright} \large

\emph{Examiner:} \\

{Dr. H.C. \textsc{van Hoof}}\\

\vspace{0.5cm}

\emph{Second reader:} \\

{N. \textsc{Butt} MSc }\\

\end{flushright}

\end{minipage}\\[2cm]




\includegraphics[width=10cm]{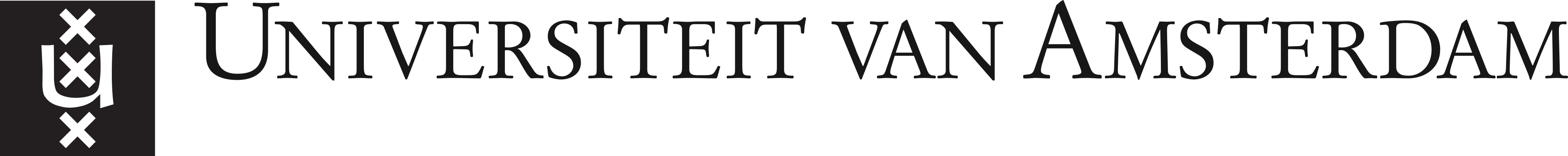}


\vfill 

\end{titlepage}

\pagenumbering{roman}

\tableofcontents

\begin{abstract}
Language models for program synthesis are usually trained and evaluated on programming competition datasets (MBPP, APPS). However, these datasets are limited in size and quality, while these language models are extremely data hungry. Additionally, the language models have a misaligned program synthesis process compared to humans. While humans iteratively develop code with the help of a compiler, most program synthesis models currently produce code in one go. To solve these issues, we introduce a bootstrapping algorithm for program synthesis, that supports teaching models how to repair. We show that bootstrapping consistently outperforms regular fine-tuning. Compared to other work, our bootstrapped model performs on par with fine-tuned models that are 68\% larger. Notably, bootstrapping with repairing also improves non-repairing performance compared to regular bootstrapping during inference. However, on our models, repairing during inference is likely inferior to simply sampling the same number of solutions. Furthermore, we find that there are issues with the example test cases in the training portion of the APPS dataset that are valuable to the community, as many repairing and reinforcement learning methods rely on them.
\end{abstract}

\pagenumbering{arabic}

\chapter{Introduction}

A longstanding goal of artificial intelligence is having machines write programs \cite{balog_deepcoder_2017}. Previously, most neural guided program synthesis approaches had to resort to using a Domain Specific Language (DSL) due to the extremely large search space of popular programming languages. This might have been limiting the adoption of these models in the real world. Recently, large language models (LLMs) have been used to circumvent the search space explosion by approaching programming as a language modeling problem. This is achieved by (pre-)training these large language models on vast amounts of publicly available code \cite{chen_evaluating_2021, li_competition-level_2022, le_coderl_2022}. These language models are then fine-tuned and evaluated on program synthesis datasets \cite{austin_program_2021, hendrycks_measuring_2021} to improve their performance on programming competition styled tasks and to show their programming performance. However, one downside of such large language models is that they require large amounts of data. Unfortunately, most program synthesis datasets are quite limited in size. Bootstrapping, which is a concept used by many reinforcement learning algorithms, might help with this issue \cite{ellis_dreamcoder_2020, haluptzok_language_2022}. \\

Bootstrapping in NLP \cite{zelikman_star_2022} and program synthesis \cite{haluptzok_language_2022, chen_improving_2023} usually entails fine-tuning on correctly generated answers by the base model. In the program synthesis world, a compiler/interpreter can be used to determine if an answer was correct or not. This is only possible if test cases are present, but most program synthesis datasets contain these (as there needs to be a way to verify the correctness of generated code). If the generated code answer was incorrect then the ground truth code answer can be used to fine-tune on. This process can be repeated for as many iterations until there is no sizeable increase in performance on the validation dataset \cite{zelikman_star_2022}. \\

Additionally, human programmers do not usually write their final programs in one go \cite{zhang_self-edit_2023}. Usually there is an iterative improvement process happening, where the programmer runs their code with an example input/test case and gets feedback from the interpreter/compiler about whether their code produces the desired result. If the result is not satisfactory, the programmer might use the interpreter/compiler feedback to improve their code and try running it again. Language models have shown to benefit from repair models/repairing techniques \cite{le_coderl_2022, zhang_self-edit_2023, chen_teaching_2023}. Some methods even include human feedback to aid in the repairing process to improve the base model \cite{chen_improving_2023}. \\

Moreover, language models have been shown to benefit from utilizing scratchpad approaches to solve certain (mathematical) problems \cite{nye_show_2021}. This is hypothesized to help due to the auto-regressive nature of language models. It might help offload work the language model would have to do internally otherwise. We argue this could be extended to program synthesis, where repairing can be seen as a scratchpad approach. A simple example would be when a language model missed an import for a package that contains a function it later ends up using. Normally, a code completion model would not be able to adjust its previous output and import the required package. However, if the model is allowed to repair its previous code predictions, it would have the ability to fix these errors where required. This is another reason why models might improve from getting a second chance, even when the feedback is not extremely informative or does not exist at all \cite{chen_teaching_2023}. \\

Therefore, this thesis aims to answer the following research question: \textbf{how can self-taught programming and repairing be used to improve program synthesis performance of language models?} Performance is measured by the estimated pass@$k$ metric, described in Section \ref{lit:pass_at_k_metric}.\\

We hypothesize the following: by bootstrapping correct answers from a coding language model, training on repaired examples while bootstrapping, and using repairing during inference, we will observe increased programming performance during inference. We expect this to be due to the increased diversity and quality of the training data, and the more comparable nature of coding with repairing to human programming/ giving the model a second chance at greedy decoding. \\

Specifically, we expect the following setup to lead to increased performance: by using a coding language model, we generate a code solution for a programming problem. Then, using the hidden test cases (during inference: the example test case), we decide whether to repair a code solution. If we decide to repair, we can show the compiler/interpreter information to the model. If at any point the model produces the correct answer, we reinforce this answer during the fine-tuning step (bootstrapping). If at the end the model turns out to have failed to produce a correct solution, we train on the ground-truth solution. The prompt we reinforce on might contain the previous solution attempt and interpreter/compiler information if repairing is taking place. We expect the model to improve from receiving:
\begin{itemize}
    \item More diverse training data, including information from the compiler/interpreter, which might be of higher quality than that of the ground-truth of the dataset.

    \item A second shot at the problem, especially when greedily decoding. Since humans use this iterative improvement process, and language models improve from having a scratchpad, it is likely combining this with bootstrapping will have positive effects.
\end{itemize}

The implications of this improvement would be numerous. First of all, if this approach of bootstrapping and repairing does indeed improve program synthesis, humanity will have better performing program synthesis models that can assist programmers in their day to day tasks, increasing their productivity \cite{ziegler_productivity_2022}. Furthermore, it is possible that with this method, smaller language models come close to the performance of much larger ones, allowing for more accessibility of models on consumer cards, faster inference times (without repairing), and less wasted electricity. If the coding model is ``self-taught" with only a few ground truth examples, there might be a smaller requirement for expensive manual annotation of the datasets. This would save a considerable amount of time and money as well. \\

There are also downsides to this approach. Repairing means that the model(s) will be asked to do inference at least twice, and this process is therefore sequential (repairing can not start before an initial code completion is given). This means it is more expensive and more time consuming than focusing on getting the right answer in one go. This method will also rely on an interpreter/compiler being present (a reasonable assumption) and test cases being present (not a guarantee, though programmers test their code all the time and sufficiently advanced language models can create their own). Bootstrapping is also a very costly approach during training time (fine-tuning the pre-trained model repeatedly), however, if this improves the performance of smaller language models that are put to use globally, then the savings in inference cost might make up for it. \\

Our main contributions are: (1) we introduce a bootstrapping algorithm for program synthesis that teaches models to repair code, (2) we demonstrate that bootstrapping consistently outperforms regular fine-tuning while achieving comparable performance to models 68\% larger, and (3) we identify significant issues with example test cases in the APPS training dataset. To support reproducibility and future research, we make our code and experimental setup available at \href{https://github.com/NoahVl/Dr-Boot}{https://github.com/NoahVl/Dr-Boot}.

\chapter{Related work}

\section{Program Synthesis Overview}
The field of program synthesis focuses on automatically creating programs that solve a specified task. It has been heralded as the ultimate abstraction in programming. To get a certain result, one can simply tell a computer what they want (through: formal specifications, input-output pairs, demonstrations, or a reference implementation), rather than how to achieve this (traditional programming) \cite{bornholt_program_2015}. \\

Throughout the years, many attempts have been made to reach this goal. The field started with pure logic based approaches \cite{balzer_15_1985, goldberg_knowledge-based_1986}, using for example heuristic search to create programs \cite{biermann_automatic_1985}. With the increased popularity of deep learning, the field moved to neurally guided (and therefore learned) search \cite{balog_deepcoder_2017, ellis_dreamcoder_2020}. However, most of these approaches used a Domain Specific Language (DSL), since regular programming languages have an enormous search space (due to the many functions/variables and the exponential growth of the search-tree). To tackle this problem, deep language modeling of regular programming languages is now used \cite{chen_evaluating_2021}. Since interacting with the interpreter/compiler can be seen as a reinforcement learning problem, these language models are regularly combined with reinforcement learning methods to improve their programming performance \cite{le_coderl_2022, shojaee_execution-based_2023, liu_rltf_2023}. \\

Recently, with language models taking over the field, natural language prompts can be used rather than needing to define a formal specification of the problem. This means that people can ask machines in natural language to write programs for them.

\section{Benchmarks and Metrics for Program Synthesis}
\subsection{Benchmarks}
\label{lit:benchmarks}
In the literature, there are several datasets that are used as benchmarks to compare program synthesis performance of newly trained language models. The most prominent ones are: APPS \cite{hendrycks_measuring_2021}, MBPP \cite{austin_program_2021}, and HumanEval \cite{chen_evaluating_2021}. These datasets are all written in a programming competition style. This means there is some text written in natural language, specifying the coding task, occasionally including expected output on some example tests. The model is then expected to produce code that gives the expected output on the hidden test cases (tests not seen by the model).

\subsection{Metrics}
To measure the performance of program synthesis programs on these datasets, metrics are needed. However, in the literature several metrics are used, the most prominent one being used inconsistently, making comparisons challenging. In this section we will therefore introduce several metrics and discuss their strengths and weaknesses.

\subsubsection{Pass@$k$}
\label{lit:pass_at_k_metric}

To evaluate program synthesis models on the benchmarks previously mentioned in \ref{lit:benchmarks}, the field of program synthesis primarily uses the ``pass@$k$" metric. This metric has originally been introduced by \citet{kulal_spoc_2019}. The $k$ in pass@$k$ stood for the number of generated programs. Per programming problem, $k$ solutions are generated, if any of those solutions pass the unit tests then the problem is considered solved. The pass@$k$ metric is then simply the fraction of solved programming problems in the dataset. \\

However, \citet{chen_evaluating_2021} claim that calculating pass@$k$ in this way can have high variance. Instead they propose a new formula for pass@$k$ estimation:

\begin{equation}
    \text {pass@}k := \underset{\text { Problems }}{\mathbb{E}}\left[1- \dfrac{\binom{n -c}{k}}{\binom{n}{k}}\right]
\end{equation}

To evaluate pass@$k$, they now generate $n$ samples per task (where $n \geq k$), count the number of correct samples $c$ which pass the unit tests (where $c \leq n$), and then plug the variables into the unbiased estimator. The authors also include a Python function to calculate this efficiently, as using the formula can give numerical instabilities \cite{chen_evaluating_2021}. \\ 

The benefit of this estimator is that the variance is indeed reduced. Many sampling combinations are now considered thanks to the binomial coefficients. However, to get a good estimate, $n$ needs to be a large number (for example: \citet{le_coderl_2022} uses $n=1000$), which requires a considerable amount of computational resources. \\

Papers that came out since \citet{chen_evaluating_2021} tend to use this new definition of pass@$k$ \cite{chen_codet_2022, le_coderl_2022, li_starcoder_2023}, however a paper that will be discussed later called ``Self-Edit: Fault-Aware Code Editor for Code Generation" \cite{zhang_coder_2022} seems to still use the old definition judging from their paper. This, together with different sampling methods: greedy or temperature sampling, and sampling hyperparameters, makes it difficult to compare results between papers.

\subsubsection{Pass@$t$}
\citet{olausson_demystifying_2023} introduce a new metric dubbed pass@$t$, this measures the pass rate of the tasks against the total number of tokens sampled from the model, enabling a fair comparison to purely sampling-based approaches \cite{olausson_demystifying_2023}. They argue that this metric is critical for repairing, as self-repair requires several dependent model invocations of non-uniform cost \cite{olausson_demystifying_2023}.

\section{Language Modeling for Program Synthesis}

\subsection{Introduction to Language Models}

In its most basic form, language models are systems that use statistical and machine learning techniques to predict the likelihood of a sequence of words. Despite the simplicity of their training objectives, large language models have shown strong generalization to unseen tasks \cite{gunasekar_textbooks_2023}. Applied to text, these language models can be used to predict the next token in an incomplete sentence. Currently, language models are ubiquitous in the field of AI. Specifically, scalable sequence prediction models are used as a general-purpose method for generation and representation in many machine learning domains \cite{chen_evaluating_2021}. Some of these domains include: natural language processing (NLP) \cite{brown_language_2020}, computer vision \cite{ramesh_hierarchical_2022, saharia_photorealistic_2022}, and music generation \cite{copet_simple_2023}. \\

Sequence prediction models for NLP are language models that have been trained to predict the next token in a sentence. This is usually done by training the model on a giant corpus of text data. In the case of language models for program synthesis, these are usually trained on large corpora of publicly available code. Most of these models rely on an adapted form of the Transformer \cite{vaswani_attention_2017} architecture.

\subsection{Language Models for Program Synthesis}
In this section, some of the models for program synthesis will be described. We will briefly cover the best performing most recent language models to investigate what makes them unique. The CodeT5 model will be explained in detail, as this is the pre-trained model that will be used for the experiments in this thesis.

\subsubsection{State of the Art Program Synthesis Models}
Due to the generative AI explosion and the success of Codex/Copilot \cite{chen_evaluating_2021} (one of the most popular commercial program synthesis models \cite{dastin_microsoft_2023}), many program synthesis language models have been developed. Currently, the best performing ones are: CodeGen 2(.5) \cite{nijkamp_codegen2_2023}, StarCoder \cite{li_starcoder_2023}, WizardCoder \cite{luo_wizardcoder_2023}, and lastly phi-1 \cite{gunasekar_textbooks_2023}. \\

\begin{figure}
    \centering
    \includegraphics[width=100mm]{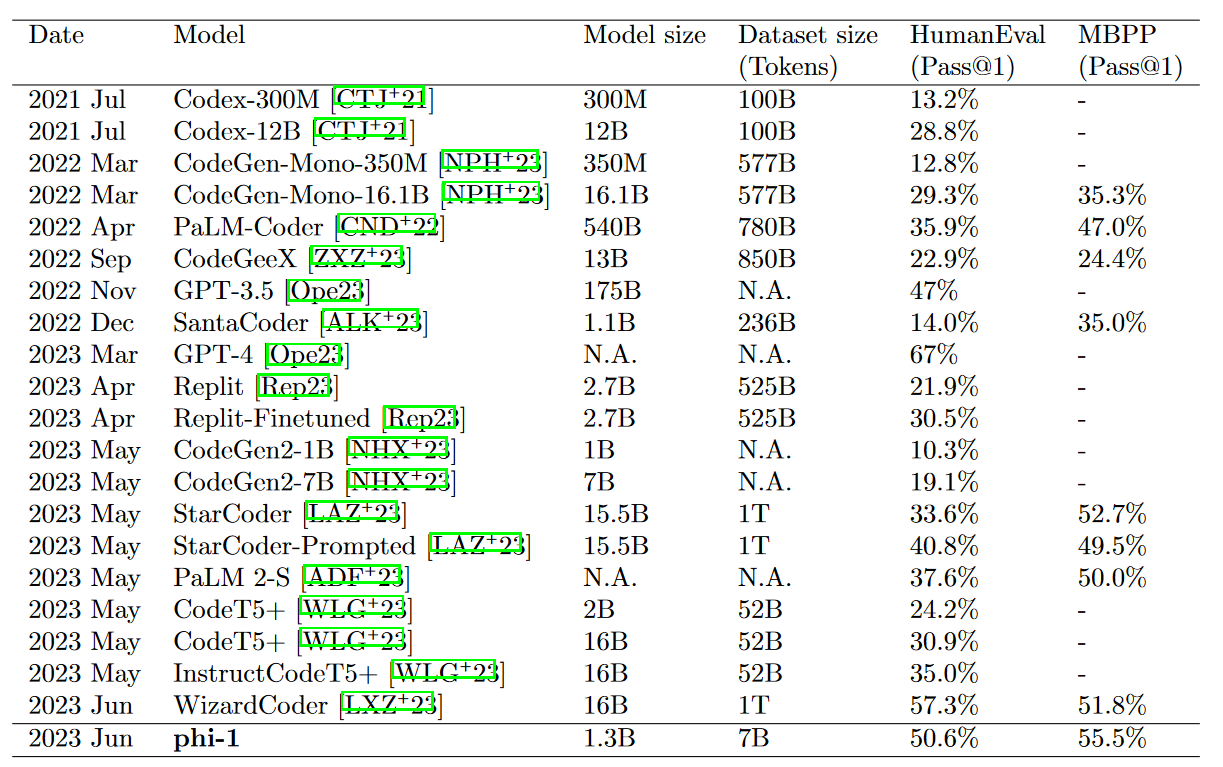}
    \caption{Table taken from \citet{gunasekar_textbooks_2023} They use self-reported scores whenever available. Despite being trained at vastly smaller scale, phi-1 outperforms competing models on HumanEval and MBPP, except for GPT-4 (also WizardCoder obtains better HumanEval but worse MBPP) \citet{gunasekar_textbooks_2023}.}
    \label{mean_top1}
\end{figure}

Figure \ref{mean_top1}, while showing the programming performance of these models, also shows the rapid advancement in the field. Many models are released each month. The way these models commonly differ is by using only some components of the Transformer \cite{vaswani_attention_2017} architecture (encoder/decoder only), using different attention mechanisms (multi-query attention \cite{shazeer_fast_2019}), different pre-training tasks \cite{wang_codet5_2021, le_coderl_2022, wang_codet5_2023}, and having been trained on different datasets. \\

Since Phi-1 \cite{gunasekar_textbooks_2023} performs best considering its size, we will investigate what makes it stand out from the other models in the next section.

\subsubsection{Textbooks Are All You Need: Phi-1}
The paper of Phi-1 \cite{gunasekar_textbooks_2023} shows how important high quality data is for these language models. They train a 1.3B parameter model which outperforms much larger ones, simply by having higher quality data. Firstly, they train a model to recognize code of high educational value, using GPT-4. They then use this classifier to train Phi-1 on only data that is deemed educationally valuable, they call this: textbook quality (6B tokens). Furthermore, GPT-3.5 is used to generate synthetic textbooks and exercises (1B tokens). The data cleanup and teacher-student training appears to improve performance substantially at a much lower training cost than other large language models.

\subsubsection{CodeT5(+)}
\label{lit:codet5}
Released a month after Codex \cite{chen_evaluating_2021} was previewed, CodeT5 is a pre-trained encoder-decoder model for code understanding and generation \cite{wang_codet5_2021}. \citet{wang_codet5_2021} pose that, while large strides have been made in applying language models to program synthesis, the approaches of them have been suboptimal. This is because most of these successes came from encoder-only models like BERT \cite{devlin_bert_2019} or decoder-only models like GPT (Generative pre-trained transformer \cite{radford_improving_2018}). Encoder models traditionally struggle with text generation tasks (as an additional decoder has to be trained which did not benefit from pre-training), while decoder-only models traditionally struggle with task understanding. \\

In addition, conventional NLP approaches regard the source code as a sequence of tokens like in natural language. This means most of the rich structural information in code is ignored, while this information can contain important code semantics. To remedy this, they propose a novel identifier-aware objective that trains the model to distinguish which tokens are identifiers (such as variable/function names) and recover them when they are masked \cite{wang_codet5_2021}. \\

The CodeT5 model has been trained with a maximum source sequence of 512 tokens \cite{wang_codet5_2021}. One can increase this context length during inference and training, however this will come at a substantial memory cost. \\

The CodeRL \cite{le_coderl_2022} paper (which will be discussed in more detail in section \ref{literature_coderl}) introduces, among other things, a new CodeT5 model. While the pre-training tasks in the original CodeT5 like Masked Span Prediction (MSP) benefit code understanding, they have a large discrepancy with program synthesis objectives \cite{le_coderl_2022}. This is why \citet{le_coderl_2022} add another pre-training task of next-token prediction (NTP) to the training of CodeT5. Using this addition to the pre-training tasks and additional datasets, they train the \texttt{Salesforce/codet5-large-ntp-py} model. This model contains 770M parameters and is therefore much smaller than Codex. Despite its size, it still manages to outperform (non fine-tuned) Codex by over 100\% with an estimated pass@1 in APPS , after fine-tuning on it \cite{le_coderl_2022}. \\

Though not used in this thesis, recently CodeT5+ was released \cite{wang_codet5_2023}. CodeT5+ is a family of encoder-decoder LLMs for code in which component modules can be flexibly combined to suit a wide range of downstream code tasks \cite{wang_codet5_2023}.

\subsection{Program Repair}
Recently many papers have been published on the topic of repairing programs \cite{chen_improving_2023, zhang_self-edit_2023, chen_teaching_2023}. This is especially relevant in the programming competition dataset space, where compiler/interpreter feedback could be made available to the model. The example tests in the problem descriptions could be used to see if there is any chance of passing the hidden tests. If the program does not pass the example tests, then it should be repaired. Repairing might improve performance over sampling more iterations as interpreter/compiler feedback adds external information, and the model gets another shot at the problem. This might be especially useful as a way to greedily decode twice.

\subsubsection{Teaching to Repair}
In fact, \citet{le_coderl_2022} use this exact approach in the CodeRL framework. They train a separate program repair model, which exploits the faulty generated programs that were originally used during RL training, as the buggy programs. The ground-truth programs were used as the target programs. Then, during inference, the programs that fail on the example test cases are each offered to the repair model together with the error type/message such that it can be repaired. This cycle can be repeated as many times as the user wants, though they find that only repairing once seems to give the best performance \cite{le_coderl_2022}. \\

\citet{zhang_self-edit_2023} take a similar approach, but instead train a small repair model (PyCodeGPT-110M) in ``Self-Edit: Fault-Aware Code Editor for Code Generation" per large coding language model to repair programs. They also show that repairing can have a large effect on the coding performance of these models. Specifically, they show that sometimes their editor can outperform having the original model generate twice the number of samples. Since the repair model is much smaller than the original models it has been trained on, this is an important finding for efficiency reasons. \\

\citet{chen_improving_2023} propose a method where humans give feedback on generated solutions that fail the unit tests. They then pass the failed generated answer and the human feedback to a trained refinement model, which attempts to repair the code. The program synthesis model is then fine-tuned on correct refinements. They suggest that learning from human-written natural language feedback is both more effective and sample-efficient than training exclusively on demonstrations for improving an LLM’s performance on code generation tasks \cite{chen_improving_2023}. \\

To the best of our knowledge, no papers have attempted to combine the bootstrapping of solutions together with repairing for multiple iterations. From the papers surveyed, it seems that they either use human feedback in between the improvement steps \cite{chen_improving_2023}, or they only train a single repairing model on faulty previously generated data without additional interactions \cite{le_coderl_2022, zhang_self-edit_2023}.

\subsubsection{Repairing through Prompting}

In ``Teaching Large Language Models to Self-Debug" \cite{chen_teaching_2023} the repairing concept is used without training. Purely by prompting Codex with repair examples and by leveraging interpreter information on MBPP, they take greedy decoding performance from 61.4\% to 69.4\% \cite{chen_teaching_2023}. Additionally, they investigate the effects of changing the amount of compiler information given to the model on performance. They find that simply informing the model the previously generated code was incorrect is similar to displaying the error/output the previous code generated (simple feedback: 68.2\%, unit test feedback: 69.4\% while greedily decoding).

\section{Reinforcement Learning in Program Synthesis}
The problem of program synthesis can be modeled as a reinforcement learning (RL) problem, especially when there are clear goals (such as passing example/hidden test cases). The coding model (agent) interacts with the environment (compiler/interpreter) and gets a reward (pass/fail/error). However, such methods require there to be example test cases in the dataset to give feedback to the model during inference.

\subsection{Actor-Critic Approaches}
\label{literature_coderl}
CodeRL \cite{le_coderl_2022} makes use of this. In their paper, the code generating LM is treated as an actor network, and they introduce a critic network that is trained to predict the functional correctness of generated programs and which provides dense feedback signals to the actor \cite{le_coderl_2022}. During inference, they introduce a new generation procedure called ``critic sampling". This allows the model to automatically regenerate programs based on feedback from example unit tests and critic scores. This approach was shown to achieve significant performance gains, outperforming many pretrained LMs of much larger sizes \cite{le_coderl_2022}. \\

Another RL approach that uses the same CodeT5 model is called PPOCoder \cite{shojaee_execution-based_2023}. They claim that approaches at the time ignore specific sequence-level features of code, such as: compilability as well as syntactic and functional correctness. They propose a training method to combine Proximal Policy Optimization (PPO) with pretrained coding models. The unique features they add during training are:
\begin{itemize}
    \item A semantic match score (based on the similarity of the proposed program and the ground-truth program). For this, they use a Data Flow Graph (DFG) of both programs.
    \item A syntactic match score. For this, they use an abstract syntax tree and again compute the similarity between the proposed program and the ground-truth.
\end{itemize} 

These features, together with compiler feedback and some training tricks result in better performance than CodeRL  \cite{le_coderl_2022} during inference (using the same critic sampling approach) in APPS (albeit very marginal) and MBPP. The method is also applied to more versatile tasks such as code completion \cite{shojaee_execution-based_2023}. \\

The downside of these actor-critic approaches is that a critic model has to be trained and used during inference. This makes training and inference more computationally expensive and requires more memory.

\subsection{Bootstrapping}
Bootstrapping has many meanings depending on the context it is used in. In the field of reinforcement learning this usually means updating a value based on some estimate(s). However, the term is also used in NLP and in program synthesis. In these cases it refers to improving a model using its own output. This is also the definition we use for bootstrapping in this thesis.

\subsubsection{Bootstrapping in NLP}
\citet{zelikman_star_2022} introduce a method called: ``STaR: Self-Taught Reasoner, Bootstrapping Reasoning With Reasoning". This method aims to improve reasoning in language models by iteratively fine-tuning the pre-trained GPT-J model on its own (fine-tuned) reasoning that led to a correct answer. They show that STaR significantly outperforms larger models on reasoning tasks such as CommensenseQA and outperforms regular fine-tuning on GSM8K (Grade School Math). \\

The bootstrapping algorithm we propose in this thesis, Algorithm \ref{algo:bootstrapping}, is inspired by the STaR \cite{zelikman_star_2022} approach but adapted to program synthesis and supports bootstrapping models to repair. In the next section we will describe how other papers apply bootstrapping to program synthesis. 

\subsubsection{Bootstrapping in program synthesis}

Bootstrapping in program synthesis seems to be rare, so we will explain the papers that do use this method in more detail. \\

In the seminal work ``DreamCoder: Growing generalizable, interpretable knowledge with wake-sleep Bayesian program learning", \citet{ellis_dreamcoder_2020} present a system that learns to solve problems by writing programs. It uses a wake-sleep approach where during the wake phase DreamCoder tries to find the best solution (program that solves a task) given the current library $L$ (the library contains abstracted functions/programs created by DreamCoder). During the sleep phase the library is updated with new programs found while awake, and the (recognition) model $Q(\rho \mid x)$ is trained to predict the best program $\rho$ given a task $x$. The dataset to train the recognition model is constructed from two pathways. Firstly, by ``fantasizing", where programs get drawn from the library and the output of these programs function as the task $x$. Secondly, by using replays. This means that tasks that have been solved during waking and their corresponding programs, will also be added to the dataset. Two types of bootstrapping are happening here. During each sleep cycle, the future library bootstraps off the concepts learned in previous cycles, growing an increasingly deep learned library \cite{ellis_dreamcoder_2020}. Simultaneously, the generative (written as P$[\rho \mid L]$) and recognition models bootstrap each other. A more specialized library will result in richer dreams for the recognition model to learn from, resulting in a better recognition model that can solve more tasks while waking, which then improves the next library while sleeping. However, the authors use a functional programming language as their DSL, which likely makes creating abstractions more tractable. It is not clear how to adapt this method to programming competition style datasets. \\

In “Language Models Can Teach Themselves to Program Better”, \citet{haluptzok_language_2022} show that by having language models create their own programming puzzles \cite{schuster_programming_2021} with solutions judged by an interpreter, and one fine-tunes them on this synthetic dataset, performance in this puzzle domain can be drastically improved. However, these puzzles and solutions are all defined in Python and might therefore not generalize well to natural language style programming competition datasets. It might also be quite difficult for language models that have been trained mostly on code to generate sound natural language prompts. \\

To the best of our knowledge, there are no papers on bootstrapping using multiple fine-tuning iterations based on interpreter feedback in programming competition datasets, despite these bootstrapping method showing promising performance. Therefore, this is a novel area of research.

\chapter{Datasets}
\label{chap:datsets}

In this chapter, the datasets that have been used will be described, explained and analyzed. While working with them, several issues came to light that might have large implications for the field as a whole. Therefore, our findings justify a chapter dedicated to only this topic.

\section{Introduction}
In the program synthesis using (large) language models literature, there are a few very popular datasets which almost all methods are evaluated on. They are called the Mostly Basic Programming Problems (MBPP) dataset \cite{austin_program_2021} and the APPS dataset \cite{hendrycks_measuring_2021}. In the experiments of this thesis, both will be used. The APPS dataset is usually used to train and evaluate the models on, while MBPP is usually used to test few-shot/zero-shot performance after training on APPS \citep{le_coderl_2022, shojaee_execution-based_2023}. This is likely due to the small size of the MBPP dataset and the much simpler problems compared to APPS. \\

\section{MBPP}
\begin{figure}[H]
    \centering
    \subfloat[\centering Task id 601.]{{\includegraphics[width=4.95cm]{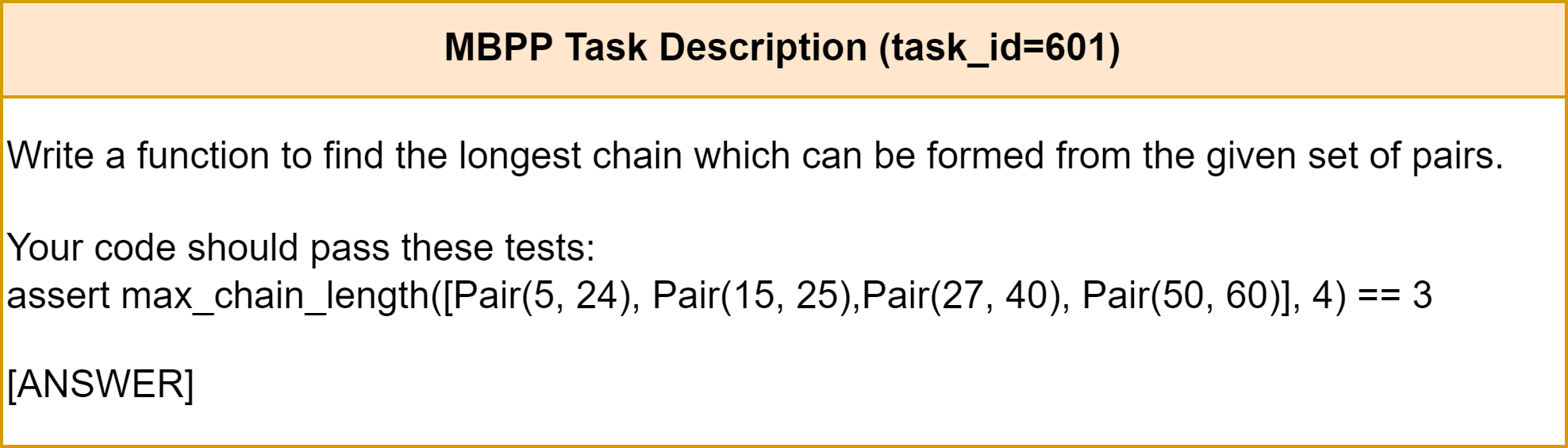} \label{fig:valid_mbpp_baselines} }}%
    \qquad
    \subfloat[\centering Task id 602.]{{\includegraphics[width=4.95cm]{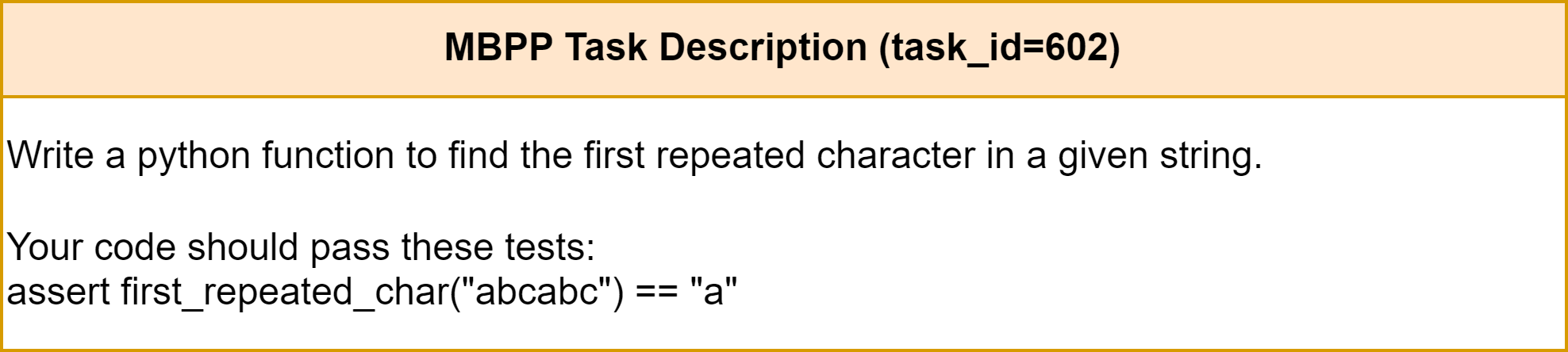} }}%
    \qquad
    \subfloat[\centering Task id 603.]{{\includegraphics[width=4.95cm]{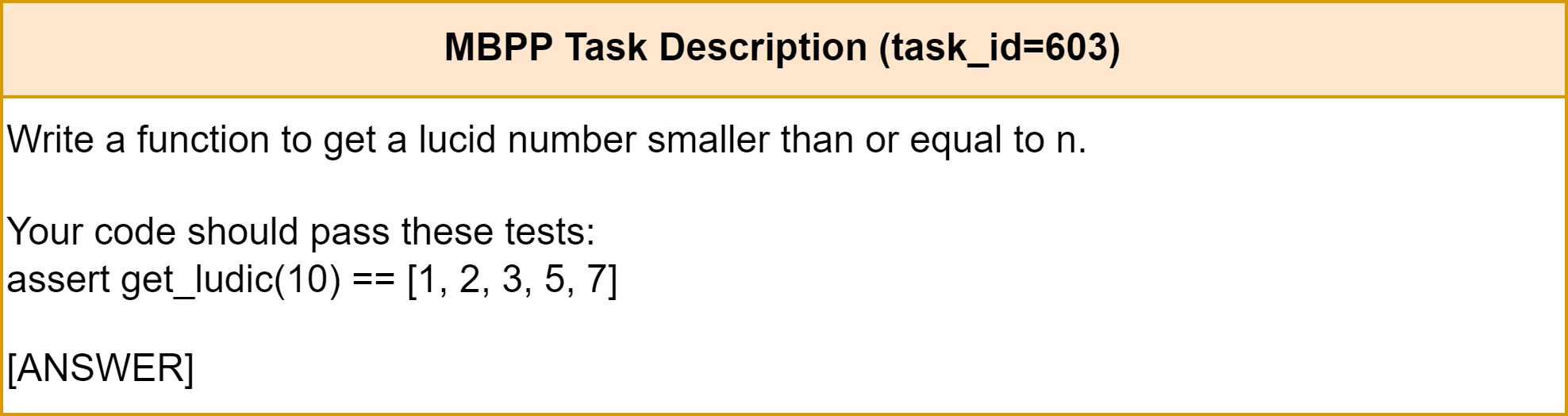}}} 
    \caption{Formatted MBPP tasks from the training dataset.}
    \label{fig:mbpp_examples}%
\end{figure}

The MBPP (Mostly Basic Programming Problems) dataset \cite{austin_program_2021} will be included for training and evaluation. In the literature this dataset is usually solely used for evaluation. The dataset contains 974 problems with a 374/90/500 split for training/validation/testing respectively. 10 problems are reserved for few-shot learning. The problems are typically short, usually one sentence of natural language descriptions each. Some examples of formatted MBPP tasks can be seen in Figure \ref{fig:mbpp_examples} \\

Each problem is accompanied by 1 correct solution (6.8 lines of code on average) and 3 unit tests, represented by asserts  \cite{le_coderl_2022}. These tests are used to see if the function is semantically correct. Unlike APPS, unit tests in MBPP are not hidden and are part of the prompt. However, we only give the first assert to the model for reasons explained in Section \ref{training_details:mbpp}.

\section{APPS}
\begin{figure}[H]
    \centering
    \includegraphics[width=170mm]{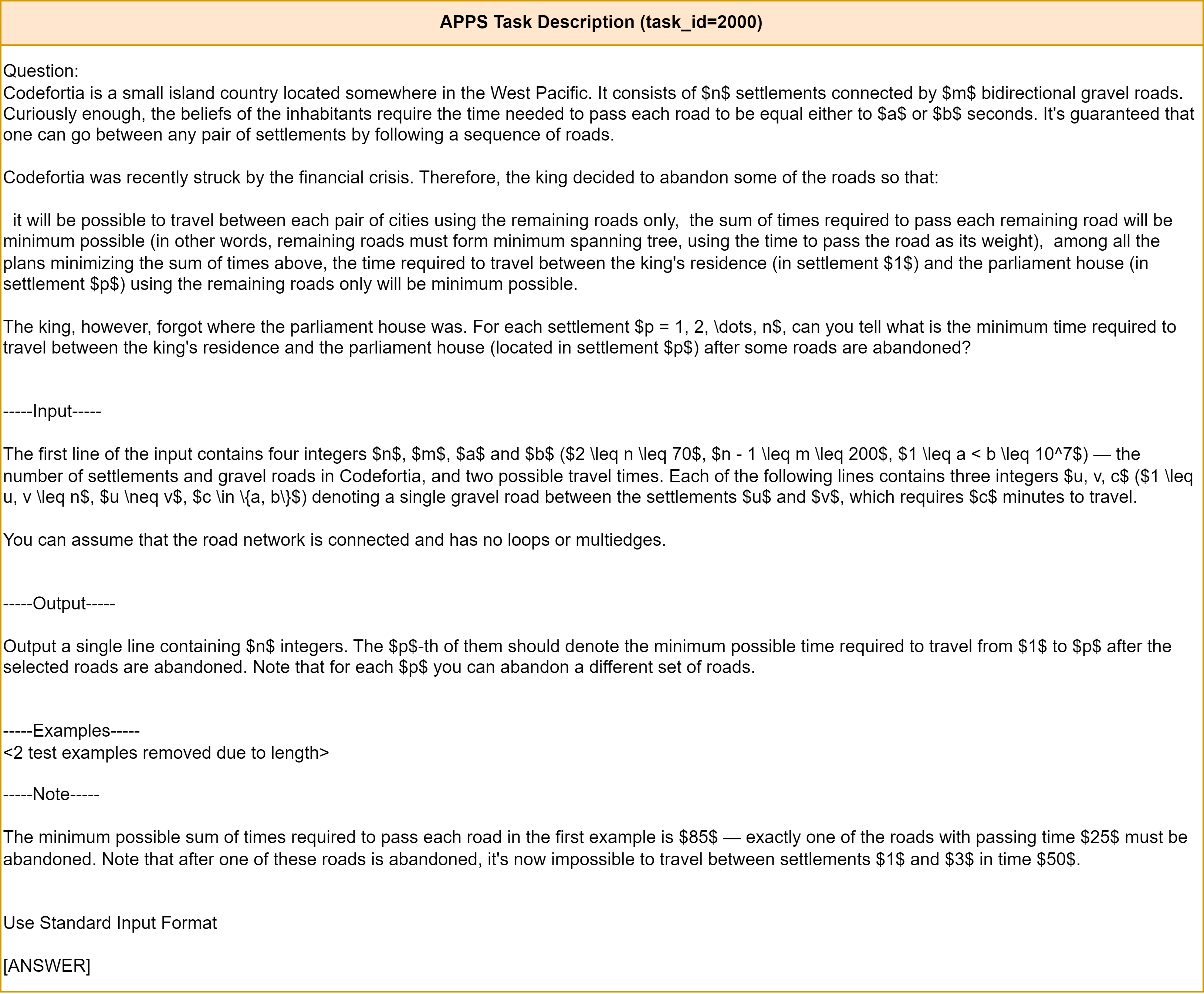}
    \caption{Example of a large APPS task description, taken from the train dataset. The problem is formatted as a story.}
    \label{fig:apps_example_task}
\end{figure}

\begin{figure}[H]%
    \centering
    \subfloat[\centering Labeled by difficulty.]{{\includegraphics[width=7cm]{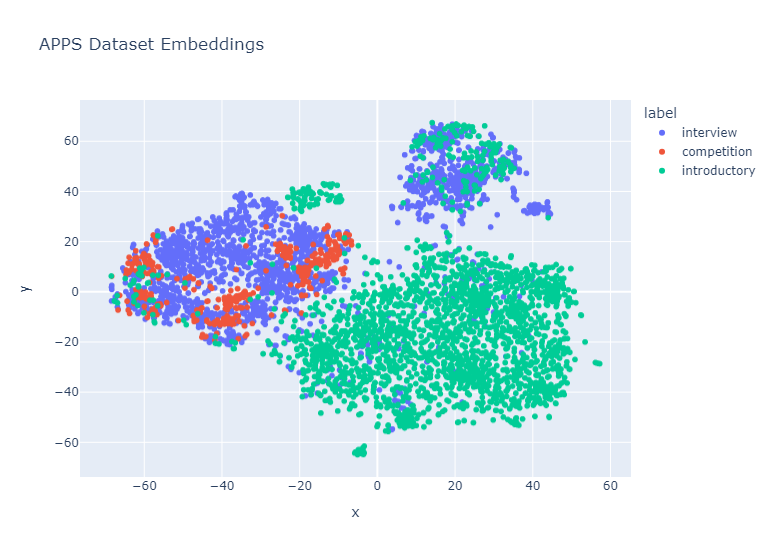} }\label{t-sne-difficulty}}%
    \qquad
    \subfloat[\centering Labeled by input type.]{{\includegraphics[width=7cm]{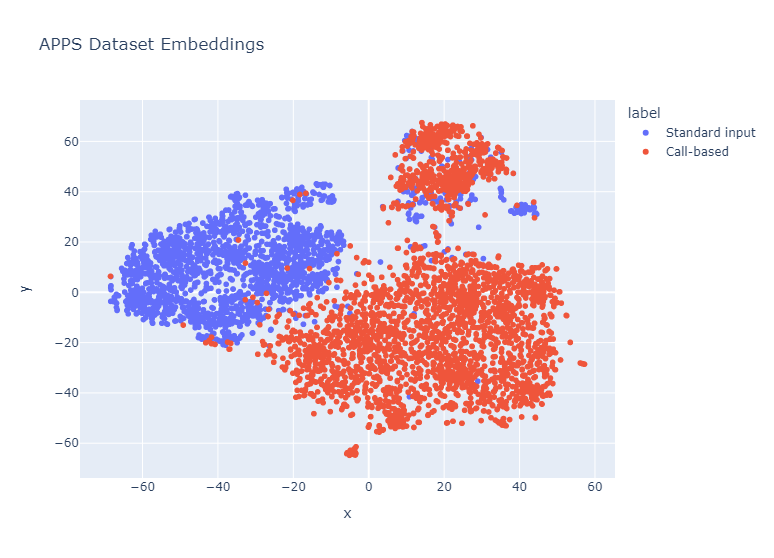} }\label{t-sne-input}}%
    \caption{t-SNE plots generated by entering the formatted APPS train tasks in the CodeT5 model and extracting the encoder's last hidden state, then averaging the token embeddings per task.}%
    \label{fig:t-sne_plots}%
\end{figure}

The APPS program synthesis benchmark \cite{hendrycks_measuring_2021} has been chosen as prior work uses it too. It has large coding problems of varying difficulties collected from multiple coding websites \cite{le_coderl_2022}. APPS consists of 10,000 coding problems with a 50-50 train-test split \cite{le_coderl_2022}. Each problem is accompanied by 23.2 correct Python programs and 21.2 unit tests on average \cite{le_coderl_2022}. The average length per problem is 293.2 words and the average length per program is 18.0 lines \cite{le_coderl_2022}. The problems are categorized in 3 levels of difficulty: Introductory, Interview, and Competition. For the distribution of these difficulties per dataset split, see Table \ref{apps_difficulties}. A t-SNE plot of the embeddings by difficulty can be seen in Figure \ref{t-sne-difficulty}.

\begin{table}[H]
\centering
\begin{tabular}{l|lll|l}
\toprule
\textbf{Dataset} & \textbf{Introductory} & \textbf{Interview} & \textbf{Competition} & \textbf{All} \\
\toprule
Train & 2639         & 2000      & 361         & 5000 \\
Test  & 1000         & 3000      & 1000        & 5000 \\
\bottomrule
\end{tabular}
\caption{Distribution of difficulty per dataset split, numbers provided by \citet{le_coderl_2022}.}
\label{apps_difficulties}
\end{table}

Problems can take two forms, they can be standard input problems or call based problems. When a problem specifies it uses the standard input, the programs need to read from \texttt{stdin}. When a problem is call based, the code verification script will attempt to call the function that is described in the prompt. The problem types are distributed non-uniformly in the dataset. In the training dataset there is a 1938/3062 split, and in the test dataset there is a 4962/38 split (standard input/call based). A t-SNE plot of the embeddings by input type can be seen in Figure \ref{t-sne-input}. \\

The same pre-processing step as \citet{hendrycks_measuring_2021} will be followed to generate the prompts from the problem task information. Additionally, we introduce a validation dataset like \citet{zhang_self-edit_2023} with 598 problems to perform early stopping. These are randomly sampled from the training dataset. \\

In the next few sections, problems that have been encountered with the dataset and related tools will be discussed.

\subsection{Large prompt size}
\begin{figure}[H]%
    \centering
    \subfloat[\centering Full-sized boxplot.]{{\includegraphics[width=7cm]{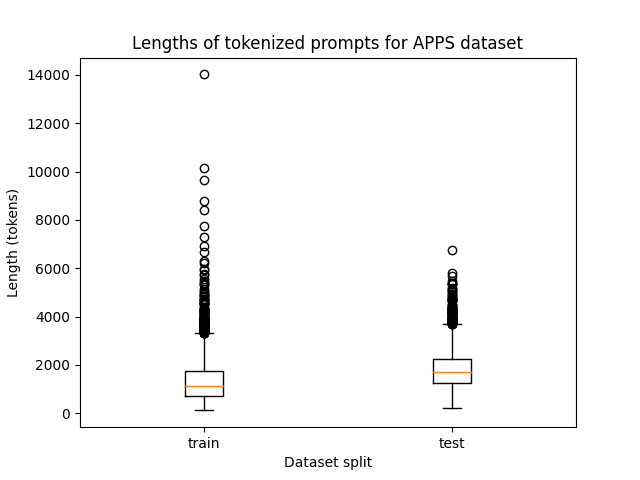} }}%
    \qquad
    \subfloat[\centering Zoomed in boxplot.]{{\includegraphics[width=7cm]{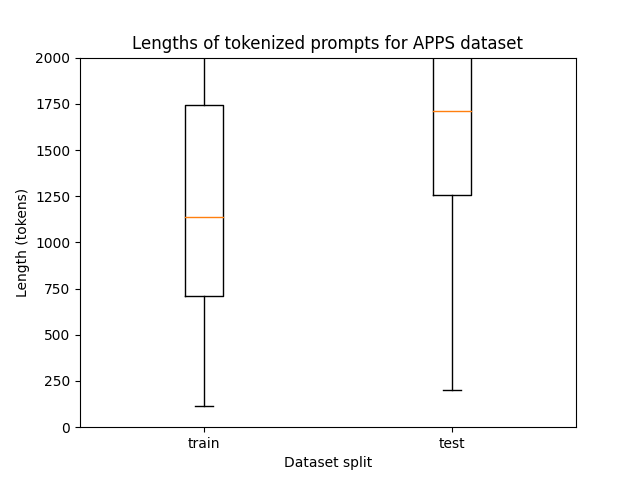} }}%
    \caption{Boxplots of the token length of each prompt in the APPS train and test dataset.}%
    \label{fig:promptsizes}%
\end{figure}

Because the maximum source sequence the CodeT5 model has been trained with is so limited, extra care has to be taken when using datasets that contain large problems. Programming problems in the APPS dataset generally contain many words, occasionally describing the programming task as a story rather than a straightforward task, see Figure \ref{fig:apps_example_task}. The number of tokens per problem were analyzed and are shown in Figure \ref{fig:promptsizes}. \\

Figure \ref{fig:promptsizes} shows that most problems fall outside the context length the CodeT5 model was trained with (512 tokens \cite{wang_codet5_2021}. This might have negative effects on the performance of the model, as it is not able to see the full task like other models can.
However, this is a limitation of the CodeT5 model and not necessarily of the dataset. Nonetheless, this model is still used on the APPS dataset by papers such as \citet{le_coderl_2022, shojaee_execution-based_2023}. Most decoder-only models that are fine-tuned or tested on the APPS dataset \cite{le_coderl_2022, shojaee_execution-based_2023} have a context length of 2048 and above \cite{brown_language_2020}.

\subsection{Example tests}
\begin{figure}[H]%
    \centering
    \subfloat[\centering Problem id 800.]{{\includegraphics[width=7cm]{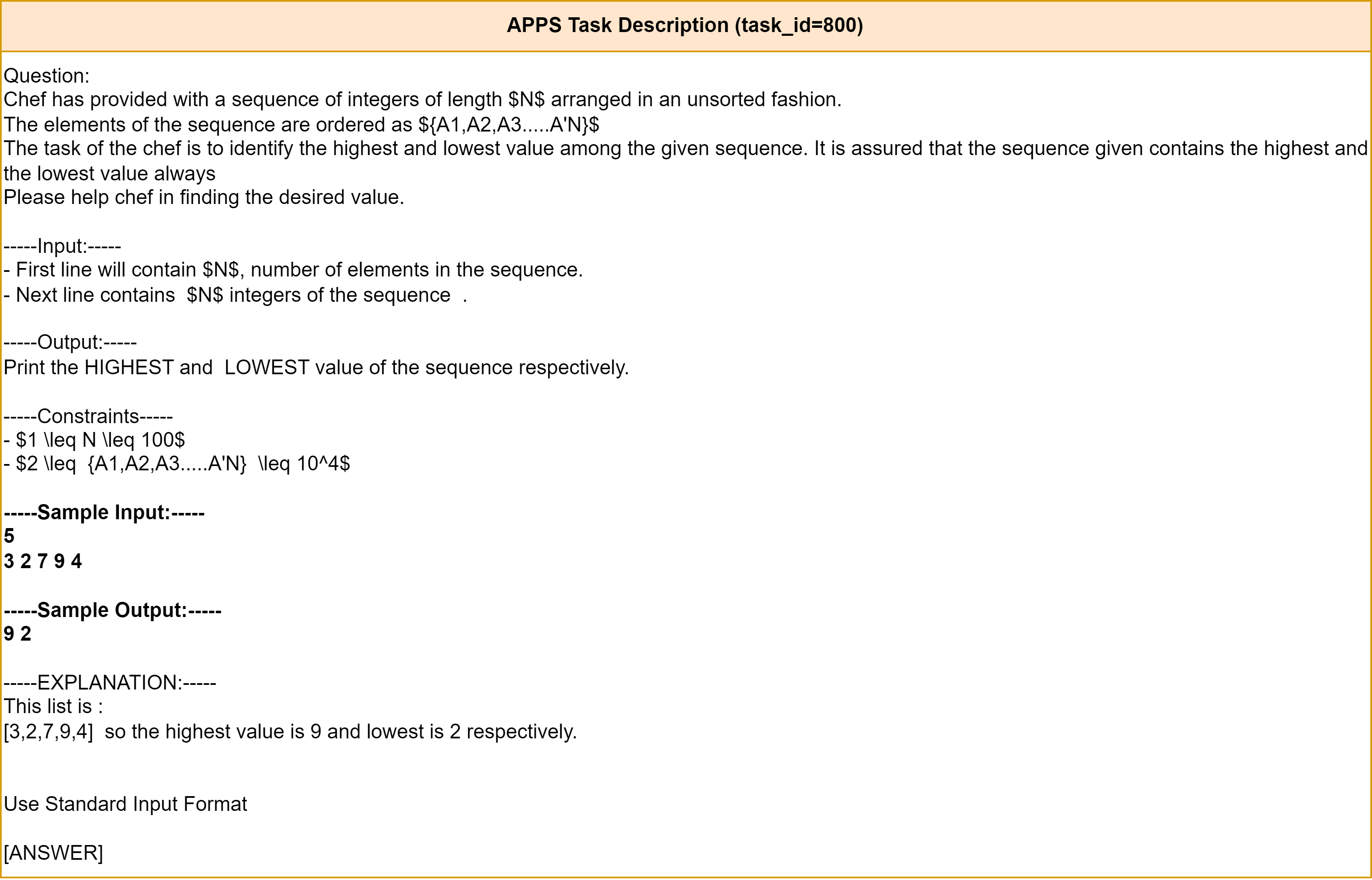} }}%
    \qquad
    \subfloat[\centering Problem id 1200.]{{\includegraphics[width=7cm]{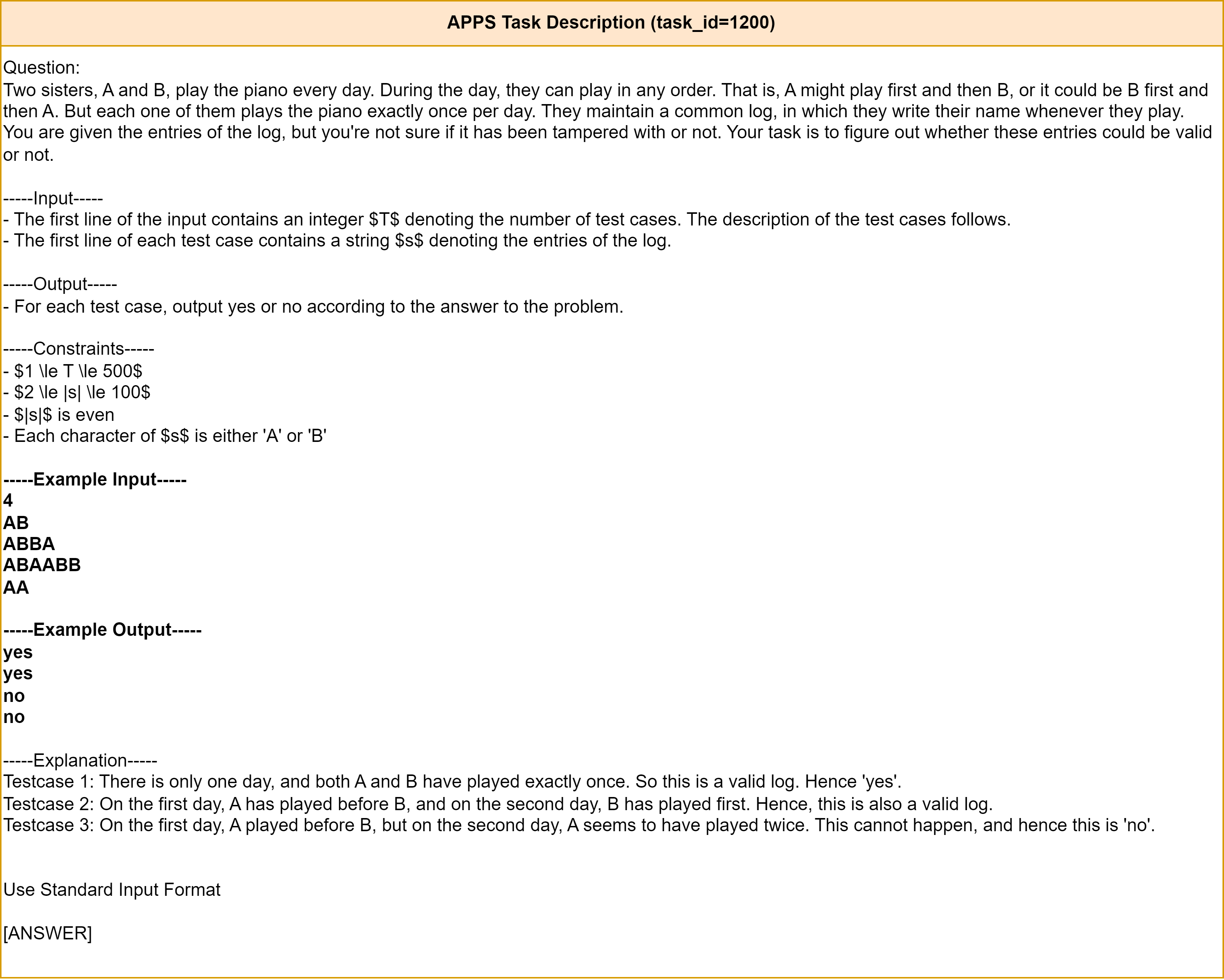} }}%
    \caption{Formatted APPS tasks from the training dataset with the example test cases marked in bold. These example test cases fully overlap with the hidden test cases.}%
    \label{fig:apps_example_tests}%
\end{figure}

APPS prompts occasionally contain example tests in the prompt, see Figure \ref{fig:apps_example_tests}. They describe the expected output for some inputs. These example tests are used by \citet{le_coderl_2022, zhang_self-edit_2023} to decide whether to repair. However, these test cases might not be as robust as initially thought as we encountered several problems with them.

\subsubsection{Noisy example test cases}
Using the extraction code and APPS program verification code by \citet{le_coderl_2022}, we evaluate the first ground-truth solution on the example test cases per problem. This results in only a 61.26\% pass@1 performance. This is in stark contrast to the 93.28\% pass@1 the first solution gets on the hidden tests of the training data. If the ground-truth solutions have trouble passing the example test cases, then the model will most likely fail as well. This would give the model a wrong training signal. \\

We observed that in some cases, the example test cases would not pass no matter what code was given. One of such cases is the example in Appendix \ref{appendix:faulty_example_test_case}. 
In this case, the repairing is done in vain, because both coding attempts pass the first example test (even though one parenthesis got changed). Coincidentally, both programs also pass the hidden test (because it is exactly the same as the first example test case). Using these example test cases would mean we are reinforcing repairing on examples that do not have to be repaired.

\subsubsection{Example tests equivalent to hidden tests}
Using this setup of extracting example test cases from the prompt, it seems that many of the test cases are equivalent to the hidden test cases in the training dataset. This means that, if a model were to overfit to these example test cases visible in the prompt, it would pass the hidden tests too. This is especially concerning for methods trying to train a bootstrapping model on this dataset, as semantically incorrect code will get reinforced. \\

It is difficult to quantify how many of these tests actually overlap (and perhaps out of scope for a thesis), because the structuring of input and output data is slightly different. However, in 89/1951 problems there is an exact match of the example and hidden unit tests. In Appendix \ref{appendix:subset_hidden_example_test_cases} there is a subset of the hidden and example test cases that were not deemed a match. As can be seen, many of them are semantically equivalent. \\

The models we trained with full access to the example tests overfit on the example test cases, this became evident from inspecting the output on the validation dataset. Therefore, it is likely that there is a considerable amount of overlap between example test cases and hidden test cases. This is why we have decided that the test examples should not be used during bootstrapping. \\

The test dataset seems to have more hidden tests and is therefore less susceptible to this total overlap of hidden tests and example tests. From a quick qualitative inspection, it seems to be usable. However, a more thorough, quantitative and critical inspection should be conducted as these example test cases are important for the repairing literature.

\chapter{Method}
This methodology should help answer the research question: "how can self-taught programming and repairing be used to improve program synthesis performance of language models?". To this end, this chapter presents a setup to bootstrap programming with repairing.

\section{Overview}
To train and evaluate our method, programming contest style datasets are used, as these are the standard benchmarks for program synthesis papers published at top conferences \cite{le_coderl_2022, shojaee_execution-based_2023}. The datasets used are described in Chapter \ref{chap:datsets}. \\

A pre-trained language model (CodeT5, see Section \ref{language_model}) will be fine-tuned and used to generate programs given prompts of programming tasks. Pre-training a language model for program synthesis from scratch would be too costly and the bootstrapping method would be very inefficient if the coding model is not already reasonably adept at producing code.

\begin{figure}[H]
    \centering
    \includegraphics[width=170mm]{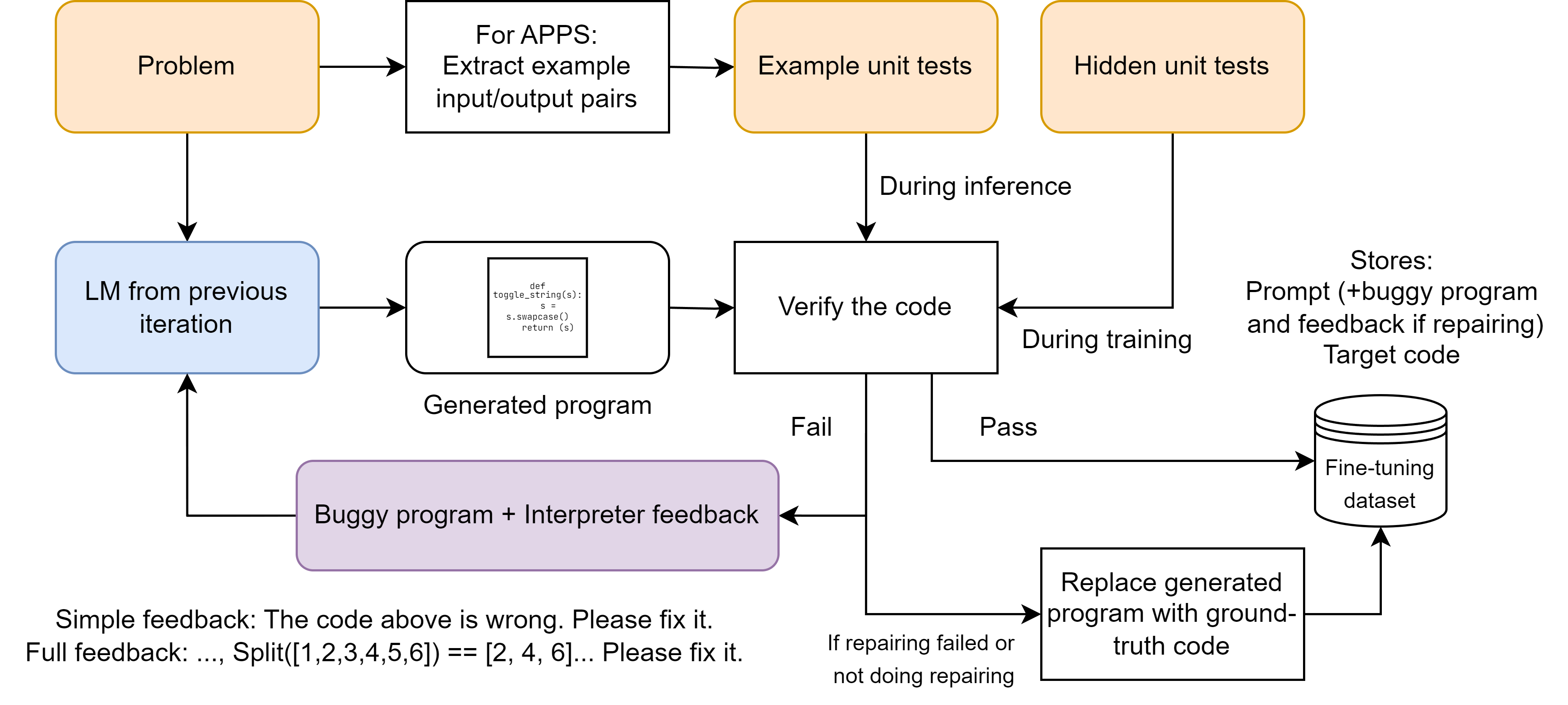}
    \caption{A visual overview of the bootstrapping algorithm (for repairing). During inference the example unit tests are used to decide whether to repair. However, for the final metrics, we evaluate on the hidden unit tests. During training, we decide whether to repair based on the hidden unit tests. The image is inspired by the Critic Sampling overview by \citet{le_coderl_2022}.}
    \label{fig:bootstrapping_visualized}
\end{figure}

Figure \ref{fig:bootstrapping_visualized} shows our proposed method: the bootstrapping algorithm for repairing. During bootstrapping, each generated program will be evaluated to see if they pass the (hidden) unit tests. If a program passes the test, it will be added to the fine-tuning dataset.

However, if the answer is incorrect, for the repairing models the next prompt will contain: the previous prompt task, the produced code, and the compiler/interpreter feedback. The repairing model is then again tasked to produce a code solution, giving it the opportunity to repair its previous response. If this response is correct, the answer is added to the fine-tuning dataset, together with the feedback containing prompt. 

If the initial code is incorrect (for the plain bootstrapping model), or the repaired code is incorrect (for the repairing models), the ground-truth solution together with the (feedback containing if repairing) prompt are added to the fine-tuning dataset. \\

Since the datasets that are used are so ubiquitous, it should be imperative to compare the results that have been achieved by this bootstrapping and repairing approach to SOTA language models and to regular fine-tuning of the baseline model. This will then show if there is any benefit of using this method to improve program synthesis performance. To this end, comparisons will be drawn with the results of the work of \citet{le_coderl_2022} and \citet{zhang_self-edit_2023}. However, because we use the APPS differently compared to the literature (see Section \ref{training_details:apps}), other methods rarely fine-tuning on MBPP, as well as to more tightly control training parameters between runs, several baselines will be trained as well. \\

A detailed description of the training process of the several baselines and methods can be found in Section \ref{method:training}. \\

\section{Language Model} \label{language_model}
\subsection{CodeT5}
As mentioned, a pre-trained model will be used as the base model in the interest of compute. Due to training/inference speed and computational budget considerations, the next token prediction (NTP) CodeT5 model specifically fine-tuned on Python code (\href{https://huggingface.co/Salesforce/codet5-large-ntp-py}{\texttt{codet5-large-ntp-py}}) will be used. This model was introduced by \citet{le_coderl_2022}, in the CodeRL paper, which will allow for a straightforward comparison with a previously state of the art method. The model has 770M parameters and uses an encoder-decoder architecture. Important to note is that the default context length it has been trained with is quite small at 512 tokens \cite{wang_codet5_2021} compared to other models. More information about the model and its training are in the previous Section \ref{lit:codet5}.

\subsection{Notation}

The model will now be referred to as: $\pi_\theta$, where $\theta$ represents the parameters of the model. Furthermore, several variables are used in this model which are defined as follows:
\begin{itemize}
    \item $x$ is the input sequence.
    \item $y$ is the target sequence.
\end{itemize}

\section{Training}
\label{method:training}

This section describes the methods of fine-tuning that have been used. First the general training method will be discussed, together with the hyperparameters used to fine-tune the models. Then, specific training details per dataset will be described. Finally, all the fine-tuning objectives of the models that have been trained are explained.

\subsection{Training method}
In this section, the general training methodology across both datasets will be displayed and justified.

\subsubsection{Bootstrapping}
We will now algorithmically define bootstrapping for the specific use-case of program synthesis, as three of the following training methods will use bootstrapping in their training objective. The definition can be found in Algorithm \ref{algo:bootstrapping}. The bootstrapping pipeline is also visualized in Figure \ref{fig:bootstrapping_visualized}.

\SetKwComment{Comment}{\# }{}
\SetKwInput{KwInput}{Input}
\SetKwInput{KwOutput}{Output}

\begin{algorithm}
\caption{Bootstrapping for program synthesis, concept and notation inspired by \citet{zelikman_star_2022}.}
\label{algo:bootstrapping}
\KwInput{\begin{itemize}
    \item $\pi_\theta$: a pre-trained LLM.
    \item programming dataset $\mathcal{D} = \{(x_i, y_i)\}^{D}_{i=1}$ (w/ few-shot prompts if used).
    \item $N$: Number of bootstrapping steps (defaults to 9).
    \item is\_plain\_bootstrapping: Whether or not to perform and train for repairing.
    \end{itemize}}
$\pi_\theta^{0} \gets \pi_\theta$ \Comment*[r]{Copy the original model}
$\mathcal{D}_r \gets \emptyset$ \;
\For{$n$ $\mathbf{in}$ $1 \ldots N$}{
$\hat{y}_i \gets \pi_\theta^{n-1}(x_i) \quad \forall i \in [1, D]$  \Comment*[r]{Generate code solutions}
$(f_i, p_i) \gets \texttt{verify\_code}(x_i, \hat{y}_i) \quad \forall i \in [1, D]$  \Comment*[r]{Interpreter feedback and pass}
$\mathcal{D}_k \gets \{ (x_i, \hat{y}_i) \mid \forall i \in [1, D] \land p_i \}$ \Comment*[r]{Save correct code}
{\eIf {$\mathrm{not}$ $\mathrm{is\_plain\_bootstrapping}$}{
        $\hat{y}_i \gets \pi_\theta^{n-1}(x_i;f_i) \quad \forall i \in [1, D] \land \neg p_i$ \Comment*[r]{Concatenate prompt and feedback}
        $(f_i, p_i) \gets \texttt{verify\_code}(x_i, \hat{y}_i) \quad \forall i \in [1, D] \land \neg p_i$; \\
        $\mathcal{D}_r \gets \{ (x_i;f_i, \hat{y}_i) \mid \forall i \in [1, D] \land p_i \}$ \Comment*[r]{Save correctly repaired code}
        } 
        {
    $f_i \gets \texttt{""} \quad \forall i \in [1, D]$ \Comment*[r]{Remove the feedback}
    }
    $\color{blue} \mathcal{D}_c \gets \{ (x_i;f_i, y_i) \mid \forall i \in [1, D] \land \neg p_i \}$
    } \Comment*[r]{Correct the faulty code}

    $\pi_\theta^n \gets $\texttt{train}$(\pi_\theta, \mathcal{D}_k \cup \mathcal{D}_r  \, { \color{blue} \cup \, \mathcal{D}_c } )$ \Comment*[r]{Fine-tune the original model on the new data}
}
\end{algorithm}

To combat overfitting, as proposed by \citet{zelikman_star_2022}, during every fine-tuning step we fine-tune on the original model. Before every bootstrapping run, the original model is also validated, showing baseline performance.

\subsubsection{Implementation and Hyperparameters}

The HuggingFace trainer is used to fine-tune the \texttt{codet5-large-ntp-py} model. To keep the comparison between all models as fair as possible, hyperparameters are kept the same across runs. The hyperparameters used during training are listed in Table \ref{table:hyperparameters}. All training and inference is performed on a single NVIDIA Tesla A100 Ampere 40 GB. For information about the versions used of Python and relevant libraries, please refer to Appendix \ref{appendix:versions}. \\

\begin{table}[htbp]
\centering
\catcode`,=\active
\def,{\char`,\allowbreak}
\renewcommand\arraystretch{1.2}
\begin{tabular}{p{7cm}<{\raggedright} p{3.5cm} p{2cm}<{\raggedright} }
  \toprule
    Training Hyperparameters           & \textbf{Values}       \\ 
    \midrule
      Batch Size                        & 6 \\
      Gradient Accumulation             & 4 \\
  \midrule
      Optimizer                 & AdamW                     \\ 
      Learning rate             & 5e-5                   \\
      Weight decay              & 0.05 \\
      Scheduler            & Polynomial \\
      Power                 & 1 \\
      Warmup Ratio          & 0.5 \\
\midrule
      Evaluate Every                    & 2 Steps \\
      Early Stopping Patience & 6 \\
      Max Epochs                      & 10 \\
\midrule
    Inference Prompt Length &      600 Tokens \\
    Inference Generation Length & 512 Tokens \\
\midrule
    Fine-tuning Prompt Length & 512 Tokens \\
    Fine-tuning Target Length & 512 Tokens \\
\midrule
    Bootstrapping Epochs            & 9 \\

  \bottomrule
\end{tabular} 
\caption{Hyperparameters used for every fine-tuning run.}
\label{table:hyperparameters} 
\end{table}

Like \citet{zhang_self-edit_2023} we use early stopping based on the validation parts of the respective datasets. This should prevent overfitting and prevent the training from taking unnecessarily long. After 6 evaluation steps of no improvement, training will stop. \\

To combat the stochasticity of the fine-tuning process and the subsequent results, several training runs will be performed per training objective such that confidence intervals can be calculated for the results section. This results in 3 models per training objective created for the MBPP dataset, and 2 models per training objective created for the APPS dataset. A larger sample size is preferred, however this was not possible due to the computationally expensive nature of the bootstrapping method and testing process. Additionally, models are only evaluated on the respective dataset that they have been trained on, unless otherwise specified.

\subsubsection{Pre-processing}
The ground-truth code is pre-processed for both datasets, using the script of \citet{le_coderl_2022}. This re-indents the code in the same way the CodeT5 model encountered code while pre-training on the GitHub dataset.

\subsubsection{Reproducibility}

Attempts have also been made to make this study fully reproducible. However, even when properly seeding the HuggingFace trainer and other libraries and using reproducible algorithms, results would still differ. Nonetheless, the code\footnote{\href{https://github.com/NoahVl/Dr-Boot}{https://github.com/NoahVl/Dr-Boot}} has been released such that results can still be somewhat reproduced, albeit with different seeds. \\

\subsection{Dataset Specific Methodology}
\subsubsection{MBPP: Training Details}
\paragraph{Few-shot examples during inference} While sampling from the MBPP models, 2 few-shot examples are added to the prompt. This is therefore also done during the bootstrapping process. These few-shot examples have been copied from \citet{chen_teaching_2023} and shortened, as the context length of the CodeT5 model is limited. The content of the few-shot examples depends on the training objective. The full prompts can be found in Appendix \ref{appendix:few_shot_examples}.

\paragraph{Fine-tuning on few-shot examples} Additionally, during fine-tuning we train on these few-shot examples as they are present in the prompts. This is also done by \citet{zelikman_star_2022, wei_finetuned_2022}, showing performance gains on their non-programming tasks while doing so. Due to this, we use left-truncation of the prompts in MBPP, as we deem the few-shot examples less important than the task at hand.

\paragraph{Preventing overfitting on test cases} \label{training_details:mbpp}Due to the bootstrapping nature of our method, we noticed that the model would start overfitting on the test cases if all 3 tests were shown to the model. This resulted in a model that would simply write out if statements for each of the test cases. While this results in impressive performance, it is not what is desired of a program synthesis model. The authors introducing MBPP also noted that programs sometimes overfit to assert statements, however in their case this happened very occasionally \cite{austin_program_2021}. 

To solve this issue, like \citet{chen_teaching_2023}, we only expose the model to the first unit test out of the 3 present in the dataset. To decide whether a solution is correct, all 3 unit tests per problem are used. During training and inference, if the model gets the first unit test correct, but not the hidden ones and is tasked to repair, a message is added saying: ``However, the code above is wrong. Please fix it.". 

\subsubsection{APPS: Training Details}
\label{training_details:apps}
\paragraph{Limited context length} Due to the limited context length the CodeT5 model was trained with, no few-shot prompts are given to the model, as in most cases the examples would get truncated. Right-truncation is used for the first pass of the problem, as this is done in \citet{le_coderl_2022}. However, if the model is tasked to repair, left-truncation is used, as we expect placing more importance on the previous output to be beneficial for repairing the previous output.

\paragraph{Test cases} Prompts are constructed in the same manner as in the APPS paper \cite{hendrycks_measuring_2021}. However, due to our findings of test example cases overlapping with the hidden tests in the training dataset, we remove them from the prompts during training. During testing, only tasks with test example cases (4954 out of 5000) are considered, otherwise there is no indication if they should be repaired or not.

\paragraph{Simulating data scarcity} We only use the first code solution of each task as ground-truth to simulate the effects of having a small dataset, to keep the runs more consistent, and to reduce computational cost during training.

\paragraph{Importing libraries} Like in \citet{le_coderl_2022}, we import a handful of libraries when testing the code the model has produced. We do this because the ground-truth does not contain imports, unlike in MBPP. This means that ModuleNotFoundError's will likely not occur.

\subsection{Trained Non-Repairing Models}
Two non-repairing training methodologies have been used to train models on both datasets to serve as baselines. These training methodologies are: regular fine-tuning and regular bootstrapping. Regular in regular bootstrapping refers to the fact that repairing is not used during training (also referred to as plain bootstrapping). The baseline methodologies will be compared to bootstrapping with repairing, to investigate if there is any performance benefit to repairing.

\subsubsection{Baseline Model, Regular Fine-tuning, $\pi_{\hat{\theta}_{brf}}$}
This is the simplest case of training and how program synthesis language models are usually trained. From the dataset, for every training example $x_i$, the model produces $y'_i$. Then using $y_i$ from the dataset as a target and $y'_i$, the cross entropy loss is computed and the model can be trained. \\

This baseline model serves two purposes. First, it will help in comparing the performance between regular finetuning and ``regular" bootstrapping (explained in the next section). Second, since it is not trained for repairing, it will help illustrate the differences in repairing performance for a model that has been trained to repair and the baselines that have not.

\subsubsection{Baseline Model, Regular Bootstrapping, $\pi_{\hat{\theta}_b}$}
For these descriptions, examples from the MBPP dataset are used for the sake of brevity. The training process would be equivalent for the APPS dataset.

\begin{itemize}
    \item $x_i$ is tokenized text representing the prompt of a programming task.
    \begin{lstlisting}[label=plain_task]
Write a python function to find the first repeated character in a given string.
        
Your code should pass these tests:
assert first_repeated_char("abcabc") == "a"

[ANSWER]
\end{lstlisting}
    \item $y_i$ is tokenized text representing a code solution to this task
\begin{lstlisting}
def first_repeated_char(str1):
  for index,c in enumerate(str1):
    if str1[:index+1].count(c) > 1:
      return c 
  return "None"
[DONE]
\end{lstlisting}

\end{itemize}
When the previous bootstrapped model produces code that passes the unit tests, that code is used to fine-tune on for the next fine-tuning iteration. If the model produces code that is incorrect, the ground-truth solution (taken from the dataset) is used to train on.
\begin{table}[H]
\centering
\begin{tabular}{|l|l|l|l|}
\hline
\textbf{Case} & Decoding & \textbf{$x_i$} & \textbf{$y_i$} \\ \hline
$\pi_{\hat{\theta}^{t-1}_b}(y'|$task) passes tests & greedy & task prompt & sampled code \\ \hline
$\pi_{\hat{\theta}^{t-1}_b}(y'|$task) fails tests & greedy & task prompt & ground-truth code \\ \hline
\end{tabular}
\label{your-label}
\end{table}

This baseline will help to illustrate the differences between a regularly bootstrapped model and one that has been trained for repairing.

\subsection{Trained Repairing Models}
\label{method:trained_repairing_models}
In this section, we propose the training methodology for two repairing approaches. These training methodologies are: bootstrapping with simple feedback and bootstrapping with full feedback. Simple feedback only informs the model that the code is incorrect, while full feedback returns errors and possibly model outputs (for MBPP). First, they are meant to show the performance difference of repairing compared to the baseline methods. Second, they will show if the amount of compiler information affects the programming performance (as has been investigated by \citet{chen_teaching_2023} on prompting alone). Unlike \citet{le_coderl_2022} and \citet{zhang_self-edit_2023} we train the same model to do both initial synthesis and code repair, we expect the parameter sharing to improve performance and make training shorter.

\subsubsection{Repair Model with Simple Feedback, $\pi_{\hat{\theta}_f}$}
The repair models are also trained using bootstrapping. When the model gets a question wrong, a feedback message is added saying: ``Feedback: The code above is wrong. Please fix it.", based on \citet{chen_teaching_2023}, and the model is tasked to repair the previous output/try again. \\

This objective only applies to the MBPP dataset, as it would be too computationally expensive to perform for APPS in conjunction with the other methods.

\begin{itemize}
    \item $x_i$ when the model answered the task correctly in one go: see previous \ref{plain_task}.
    
    $x_i$ when the model answered the task incorrectly initially:
    \begin{lstlisting}
Write a function to toggle characters case in a string.
        
Your code should pass these tests:
assert toggle_string("Python")==("pYTHON")

[ANSWER]
def toggle_string(s):
    s = s.lower()
    return (s)
[DONE]

Feedback: The code above is wrong. Please fix it.

[ANSWER]
\end{lstlisting}
Thus $x_i$ is now: the (previous) prompt, the previous incorrect answer, and a simple feedback message (which is always saying the code is incorrect).

    \item $y_i$ is tokenized text representing a code solution to this task

    \begin{lstlisting}
def toggle_string(s):
    s = s.swapcase()
    return (s)
[DONE]
\end{lstlisting}

This model is trained such that the impact of feedback quality on repairing performance can be investigated. A model might not necessarily influence its output based on verbose interpreter/compiler information. Simply giving it a second chance with the previous output could be enough for good repairing performance.

\end{itemize}
The training cases:
\begin{itemize}
    \item The model gets the code \textbf{correct in one go}. Finetune on: task prompt, generated code.
    \item The model gets the code incorrect on its first try. The model \textbf{repairs the code correctly}. Fine-tune on: task prompt + old code + simple feedback, correctly repaired code.
    \item The model gets the code incorrect on its first try. The model \textbf{repairs the code incorrectly}. Fine-tune on: task prompt + old code + simple feedback, ground-truth code.
\end{itemize}

\begin{table}[H]
\centering
\resizebox{\columnwidth}{!}{
\begin{tabular}{|l|l|l|l|}
\hline
\textbf{Case} & Decoding & \textbf{$x_i$} & \textbf{$y_i$} \\ \hline
$\pi_{\hat{\theta}^{t-1}_f}(y'|$task) passes tests & greedy & task & generated code \\ \hline
$\pi_{\hat{\theta}^{t-1}_f}(y'|$task+code+feedback) passes tests & greedy & task+code+feedback & 2nd generated code \\ \hline
$\pi_{\hat{\theta}^{t-1}_f}(y'|$task+code+feedback) fails tests & greedy & task+code+feedback & ground-truth code \\ \hline
\end{tabular}
}
\label{your-label}
\end{table}

\subsubsection{Repair Model with Execution Information, $\pi_{\hat{\theta}_r}$}
This training method is equivalent to the previous simple repair model, however now the interpreter information will be displayed to the model. This might aid the model in repairing, as humans use interpreter/compiler output to diagnose bugs \cite{zhang_self-edit_2023}. The execution information might contain Python errors such as SyntaxError, IndentationError, TypeError or information about the function being terminated due to timeout. If no errors occurred, the feedback message will contain the expected output and the produced output by the model generated code. For APPS, because during training and validation the example tests are omitted (due to them overlapping with the hidden tests), we do not return information about output and expected output (even during testing). Rather, if the outputs do not match we return the following message: ``Feedback: OutputMismatchError: The code does not pass the test. Please fix it.".

\begin{itemize}
    \item $x_i$ when the model answered the task correctly in one go: see previous \ref{plain_task}.
    
    $x_i$ in the case the model got the task wrong the first time:
    \begin{lstlisting}
Write a python function to find odd numbers from a mixed list.
        
Your code should pass these tests:
assert Split([1,2,3,4,5,6]) == [1,3,5]

[ANSWER]
def Split(nums):
  odd_nums = []
  for i in nums:
    if i % 2 == 0:
      odd_nums.append(i)
  return odd_nums
[DONE]

Feedback: With the above function, Split([1,2,3,4,5,6]) == [2, 4, 6]. The assertion is "assert Split([1,2,3,4,5,6]) == [1,3,5]". So the code does not pass the assertion. Please fix it.

[ANSWER]
\end{lstlisting}
    
    \item $y_i$ is tokenized text representing a code solution to this task

\begin{lstlisting}
def Split(nums):
  odd_nums = []
  for i in nums:
    if i % 2 == 1:
      odd_nums.append(i)
  return odd_nums
[DONE]
\end{lstlisting}
    
\end{itemize}
The training cases are the same as previous simple feedback training method, however now the feedback contains interpreter information such as errors or produced output (for MBPP).
\begin{itemize}
    \item The model gets the code \textbf{correct in one go}. Fine-tune on: task prompt, generated code.
    \item The model gets the code incorrect on its first try. The model \textbf{repairs the code correctly}. Fine-tune on: task prompt + old code + interpreter feedback, correctly repaired code.
    \item The model gets the code incorrect on its first try. The model \textbf{repairs the code incorrectly}. Fine-tune on: task prompt + old code + interpreter feedback, ground-truth code.
\end{itemize}

\begin{table}[H]
\centering
\resizebox{\columnwidth}{!}{
\begin{tabular}{|l|l|l|l|}
\hline
\textbf{Case} & Decoding & \textbf{$x_i$} & \textbf{$y_i$} \\ \hline
$\pi_{\hat{\theta}^{t-1}_r}(y'|$task) passes & greedy & task & generated code \\ \hline
$\pi_{\hat{\theta}^{t-1}_r}(y'|$task+code+int. feedback) passes & greedy & task+code+int. feedback & 2nd generated code \\ \hline
$\pi_{\hat{\theta}^{t-1}_r}(y'|$task+code+int. feedback) fails & greedy & task+code+int. feedback & ground-truth code \\ \hline
\end{tabular}
}
\label{your-label}
\end{table}

\chapter{Experiments}
In this chapter, we will first discuss the importance of proper sampling approaches 
 for experimenting. Then, we will describe the experiments that were run using these sampling methods. The results are analyzed and discussed. Finally, we will discuss some of the limitations of our work, answer the research question, and recommend steps for future work.

\section{Sampling methods}
To generate a code solution with a language model, one has to decide how to sample the output. This is especially important when introducing repairing, as the repairing models get two shots at the problem (when greedily decoding). It is therefore important to design the experiments such that they make the comparisons with non-repairing models as fair as possible.

\subsection{Baseline Models, $\pi_{\hat{\theta}_{brf}}$ and $\pi_{\hat{\theta}_b}$}
Because the baseline models have not been trained to repair, they might not benefit as much from trying to repair by greedily decoding twice. Hence, we propose different sampling methods, which we call baselines, to make the comparison with repairing models (that decode twice) fairer.

\begin{itemize}
    \item Greedy decoding (Baseline 1): to get a baseline comparison result to compare with the initial pass of the repairing models.
    \item Beam search decoding (Baseline 2): a beam of 2 could be used as a way to still do greedy decoding while getting two shots at the problem. However, the beams will likely be very similar to each other. This will return 2 solutions that will be tested.
    \item Temperature sampling (Baseline 3): this is a standard way of generating multiple outputs from (program synthesis) language models. In this case, 2 solutions would be sampled and tested as well.
\end{itemize}

Additionally, we can treat the baseline models as if they were repairing models, by prompting them to repair in the same way as the repairing models.

\subsection{Repair Models, $\pi_{\hat{\theta}_f}$ and $\pi_{\hat{\theta}_r}$}
To test the greedy performance of the repairing models, we can simply decode in the same way as during bootstrapping, described in Section \ref{method:trained_repairing_models}.

However, to compare to \citet{le_coderl_2022} and \citet{zhang_self-edit_2023} temperature sampling will also be used for the repair models.

\section{Validation and Testing}
In this section we list the experiments that have been conducted during validation and testing. The validation results dictate which models we will test.

\subsection{Overview}
\subsubsection{Selecting the best model for testing}
Before testing, for every training objective run, the best bootstrapped model is selected (from the 9 bootstrapping models trained per run, the best is selected). We quantify the best models by looking at the greedy pass@1 performance on the validation dataset. The repairing (edit) pass@1 performance, if repairing was part of the training objective, is used as a tie breaker. For the regularly fine-tuned models, there is only one model, so this will be selected as the best model.

\subsubsection{Pass@$k$ estimation for testing}
Then, during testing, the best models will be subjected to temperature sampling (with $t=0.8$), generating $n=10$ solutions per problem. This is done to directly compare to the results of \citet{zhang_self-edit_2023} for the APPS dataset. From these 10 generated samples, the pass@1, pass@2, pass@5, pass@10 will be calculated. For the repair models the incorrect samples (judged by the example tests) will get one shot at being repaired (as in \cite{zhang_self-edit_2023}). The non-repairing models will also be evaluated on repairing, to see if specifically training for repairing actually gives an advantage, or if these models are inherently able to repair (without needing training). Greedy decoding is also used to compare the trained models with less variance.
\\

The sampling results will also be compared to the results obtained in the CodeRL paper \cite{le_coderl_2022}, however they use different sampling techniques, a different definition of pass@$k$ (than \cite{zhang_self-edit_2023}) and have reportedly trained on the full APPS dataset. \\

Because \citet{zhang_self-edit_2023} uses the non-estimated definition of pass@$k$, when comparing with their results we will do the same. This means we will make a subselection of the 10 generated samples to get the sampled pass@$k$. Per table we will make clear what definition of pass@$k$ (estimated or sampled) we are using.

\subsubsection{Edit pass@$k$ calculation}
We define edit pass@$k$ to be same as pass@$k$ but having had one additional chance to repair the incorrect programs. This means that ``edit pass@$k$" should be compared to pass@$2k$, as maximally $2k$ solutions were generated. The edit pass@$k$ metric will be shaded gray in tables of the results section as has been done by \citet{zhang_self-edit_2023}.

\subsection{Definitions}
Definitions:
\begin{itemize}
    \item $T$ = task description
    \item $C_o$ = old code
    \item $F_f$ = simple feedback
    \item $F_r$ = interpreter feedback
\end{itemize}

\subsection{MBPP}
On the MBPP dataset, the following experiments will be conducted on the validation and testing dataset:\\

\begin{table}[H]
\resizebox{\columnwidth}{!}{
\begin{tabular}{|l|l|l|l|}
\hline
\textbf{Experiment Name} & \textbf{MBPP Model(s)} & \textbf{Decoding Strategy} & \textbf{Model Inputs $x_j$} \\ \hline
Baseline 1 & $\pi_{\hat{\theta}_b}$, $\pi_{\hat{\theta}_{brf}}$ & greedy decoding & $T$ \\ \hline
Baseline 2 (validation) & $\pi_{\hat{\theta}_b}$ & beam search, $b=2$ & $T$ \\ \hline
Baseline 3 & $\pi_{\hat{\theta}_b}$, $\pi_{\hat{\theta}_{brf}}$ & temperature sampling, $t=0.8$ & $T$ \\ \hline
Simple Feedback & $\pi_{\hat{\theta}_b}$, $\pi_{\hat{\theta}_f}$, $\pi_{\hat{\theta}_r}$ & greedy decoding & $T$, $T$+$C_o$+$F_f$ \\ \hline
Full Feedback & $\pi_{\hat{\theta}_b}, \pi_{\hat{\theta}_f}, \pi_{\hat{\theta}_r}$, $\pi_{\hat{\theta}_{brf}}$ & greedy decoding & $T$, $T$+$C_o$+$F_r$ \\ \hline
\end{tabular}
}
\caption{Experiments on the MBPP dataset. $T$ is few-shot prompted according to the prompts in Appendix \ref{appendix:few_shot_examples}.}
\label{your-label}
\end{table}

For each training objective, 3 models were trained, giving 3 results per experiment.

\subsection{APPS}
On the APPS dataset, the following experiments will be conducted on the validation and testing dataset:

\begin{table}[H]
\resizebox{\columnwidth}{!}{
\begin{tabular}{|l|l|l|l|}
\hline
\textbf{Experiment Name} & \textbf{APPS  Model} & \textbf{Decoding Strategy} & \textbf{Model Inputs $x_j$} \\ \hline
Baseline 1 & $\pi_{\hat{\theta}_b}$, $\pi_{\hat{\theta}_{brf}}$ & greedy decoding & $T$ \\ \hline
Baseline 3 & $\pi_{\hat{\theta}_b}$ & temperature sampling, $t=0.8$ & $T$ \\ \hline
Full Feedback & $\pi_{\hat{\theta}_r}$, $\pi_{\hat{\theta}_b}$, $\pi_{\hat{\theta}_{brf}}$ & greedy decoding & $T$, $T$+$C_o$+$F_r$ \\ \hline
\end{tabular}
}
\caption{Experiments on the APPS dataset.}
\label{your-label}
\end{table}

Fewer experiments were conducted on the larger APPS dataset due to a limited computational budget. Additionally, for each training objective only 2 models have been trained, resulting in 2 samples per experiment.

\newpage

\section{Results}
In this section the results of the experiments will be shown and analyzed. The validation section shows the greedy decoding results on the validation dataset per bootstrapping step and the baseline results. The testing section shows the programming performance of the best models (selected from the validation performance), both greedy and with temperature sampling. A subset of qualitative results are shown in Appendix \ref{appendix:generated_code}.

\subsection{Validation}
Every bootstrapping step the models are evaluated on the validation dataset. The initial pre-trained model is also evaluated at bootstrapping step 0. For MBPP this shows programming performance with different prompts. For APPS the 0th bootstrap step results should all be the same, as no few-shot prompts are used there due to context length limitations.

\subsubsection{MBPP}
Figure \ref{fig:valid_mbpp_baselines} shows all the baselines for plain bootstrapping. It can be seen that Baseline 2 (beam search with 2 beams) on average performs best on this dataset. Baseline 2 has the highest performing spikes and is predominately the best performing experiment through time compared to the other baselines. \\

\begin{figure}[H]
    \centering
    \subfloat[\centering Plain bootstrapping baselines.]{{\includegraphics[width=4.95cm]{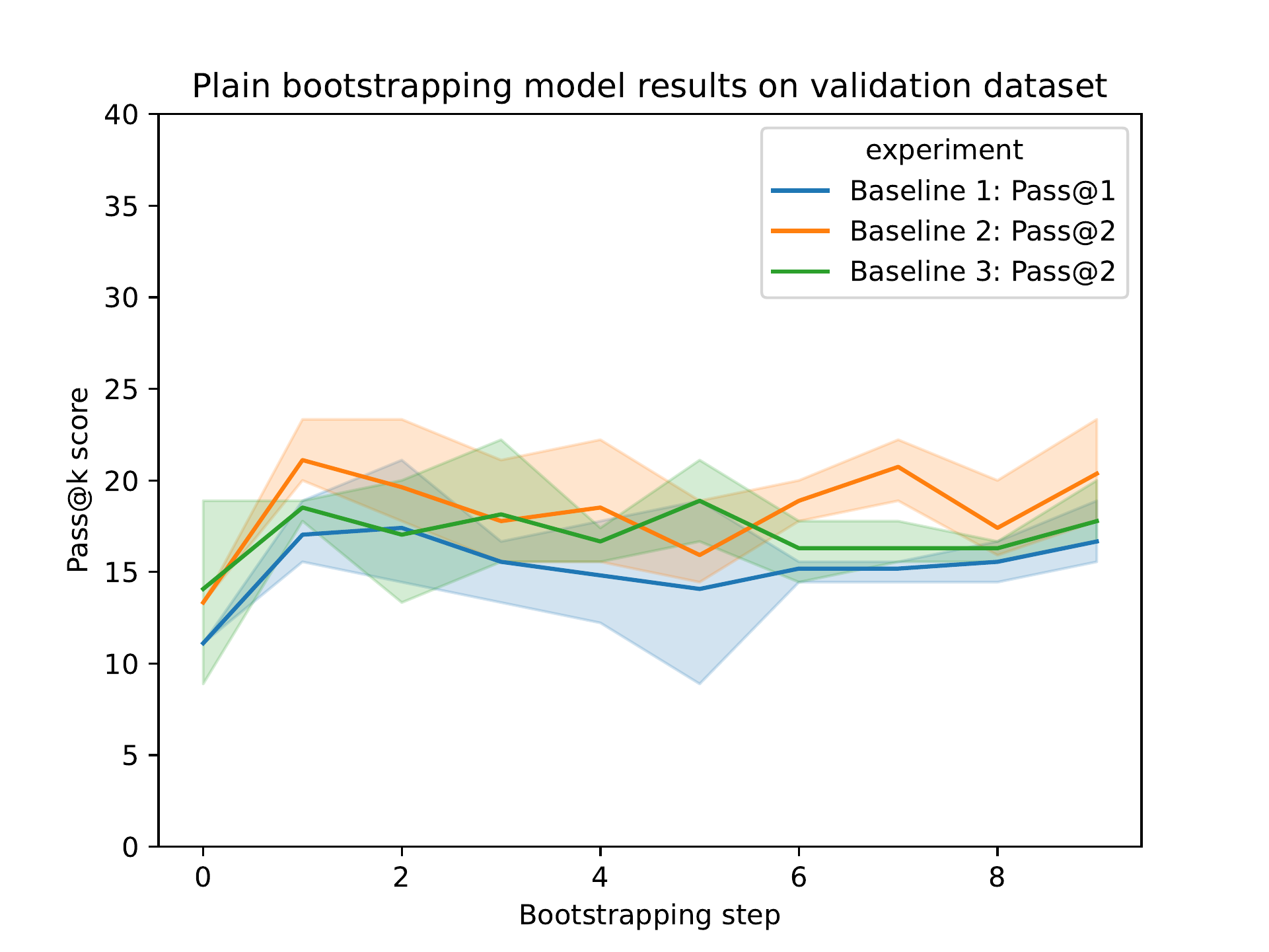} \label{fig:valid_mbpp_baselines} }}%
    \qquad
    \subfloat[\centering Non-repairing results of the 3 different training objectives.]{{\includegraphics[width=4.95cm]{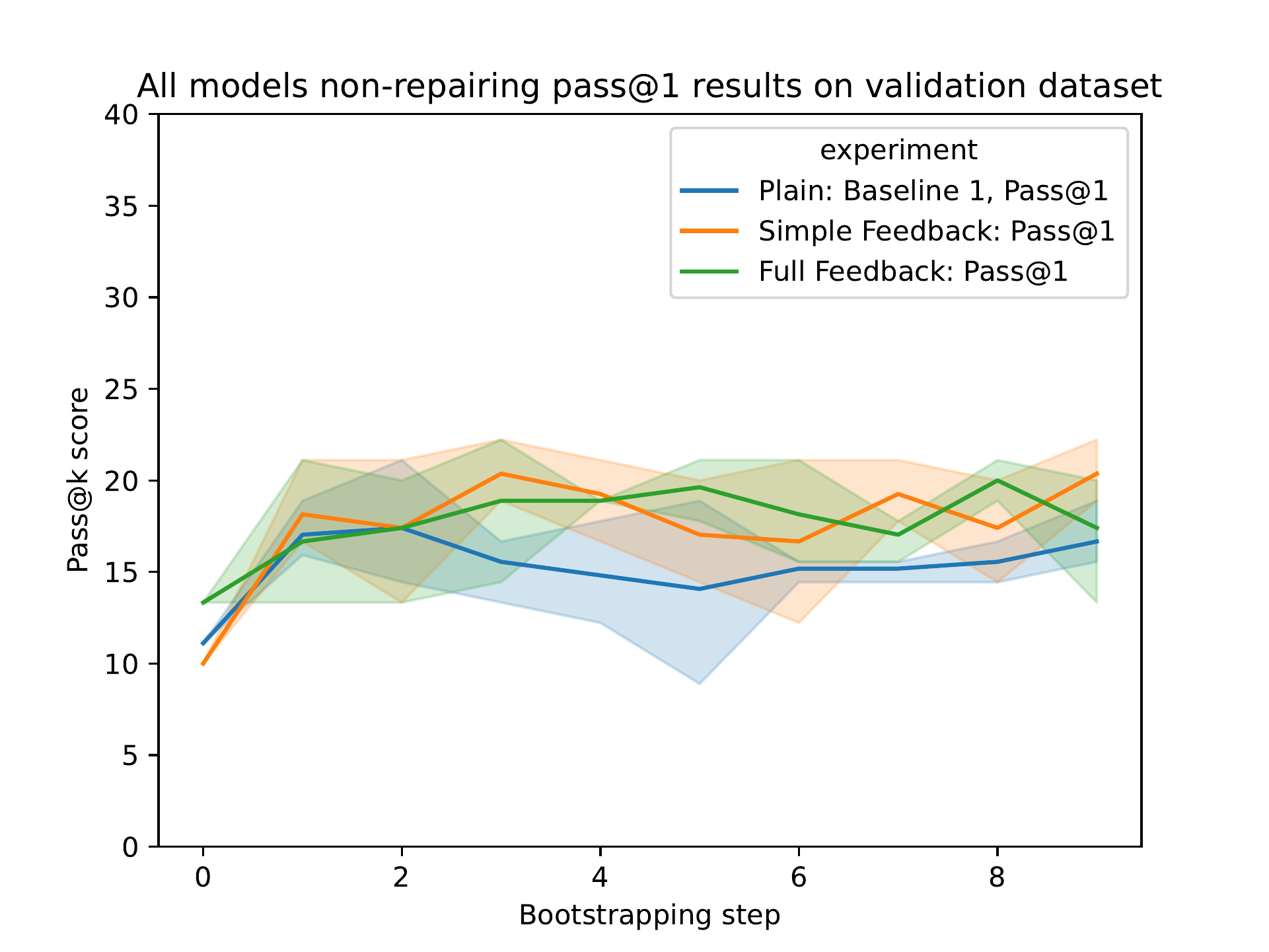} \label{fig:valid_mbpp_non_repairing_results} }}%
    \qquad
    \subfloat[\centering Repairing results of models trained to repair.]{{\includegraphics[width=4.95cm]{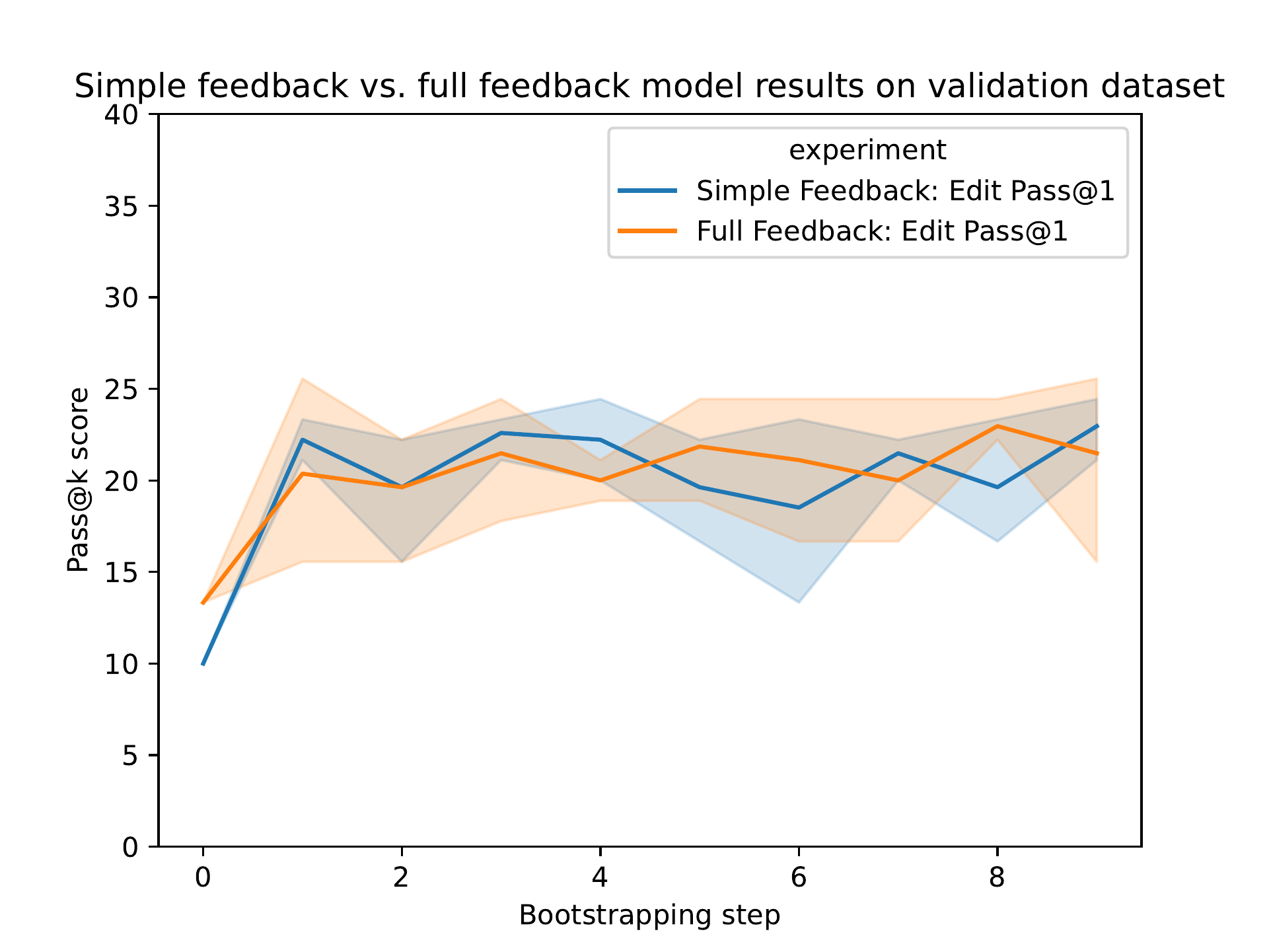} 
    \label{fig:valid_mbpp_repairing}}} 
    \caption{Results of several experiments conducted while validating. The samples come from 3 separate models trained with their respective training objective. The lines represents the average result and the shaded area represents the 95\% confidence interval.}
    \label{fig:mbpp_validation_results}%
\end{figure}

In Figure \ref{fig:valid_mbpp_non_repairing_results} the non-repairing performance of the different models through time are shown. \label{results:prompting_influences_pre_trained_model}Notably, the 3 training objectives produce a different pass@1 performance at bootstrapping step 0 (validating the pre-trained model with their respective few-shot examples). This shows that the few-shot example prompts affect the inference performance of the pre-trained model, before training. Additionally, the greedy non-repairing performance of the repairing models seems to consistently outperform the greedy pass@1 performance of the plain bootstrapping model. \\

Figure \ref{fig:valid_mbpp_repairing} shows the repairing performance of the repairing models compared to each other. The repairing performance appears to be fairly comparable through time, with the full feedback model having the highest spike. \\

\begin{figure}[H]
    \centering
    \subfloat[\centering Plain Baseline 2 and simple feedback results.]{{\includegraphics[width=8cm]{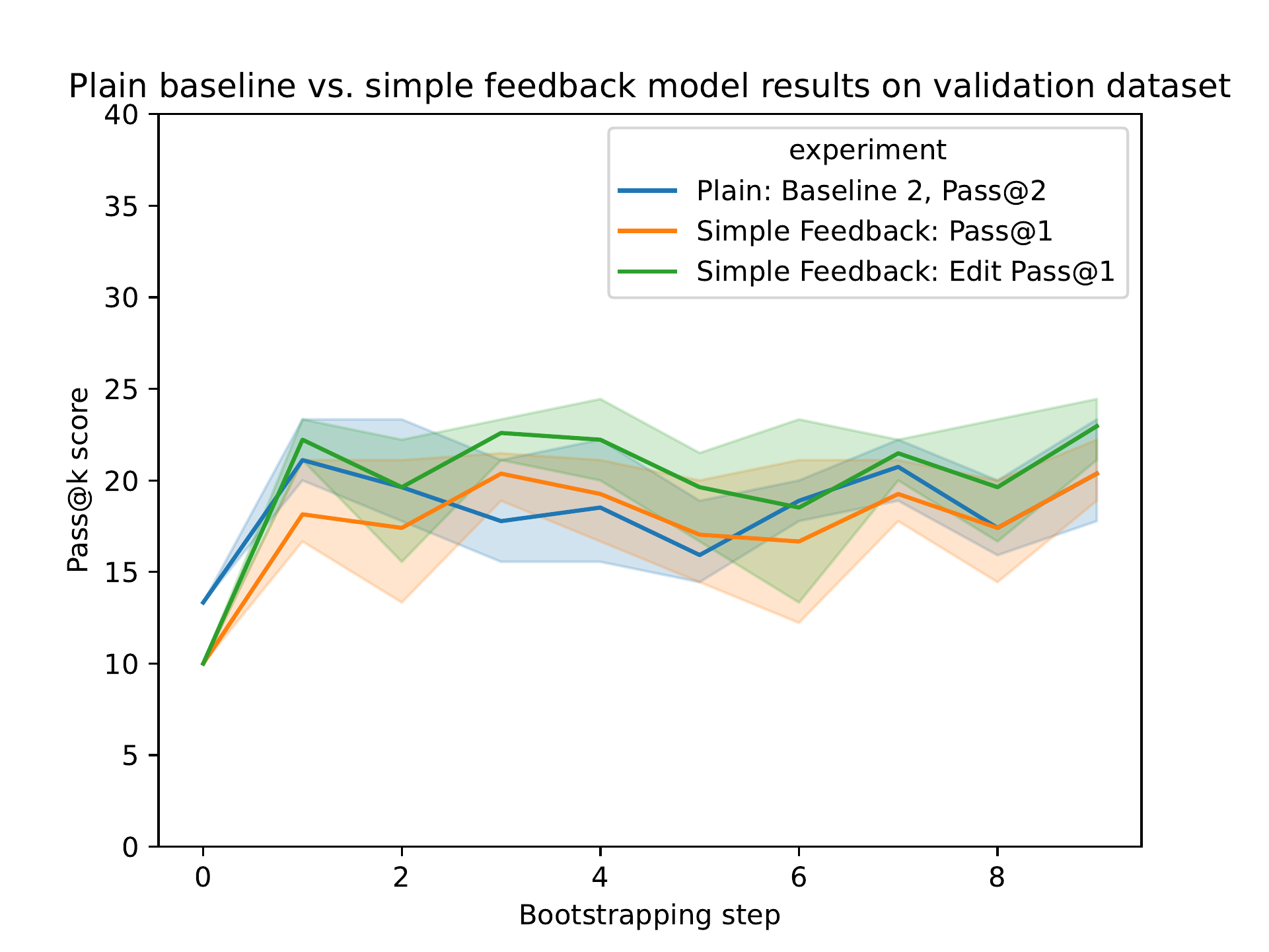}\label{fig:mbpp_simple_vs_baseline2}}}%
    \qquad
    \subfloat[\centering Plain Baseline 2 and full feedback results.]{{\includegraphics[width=8cm]{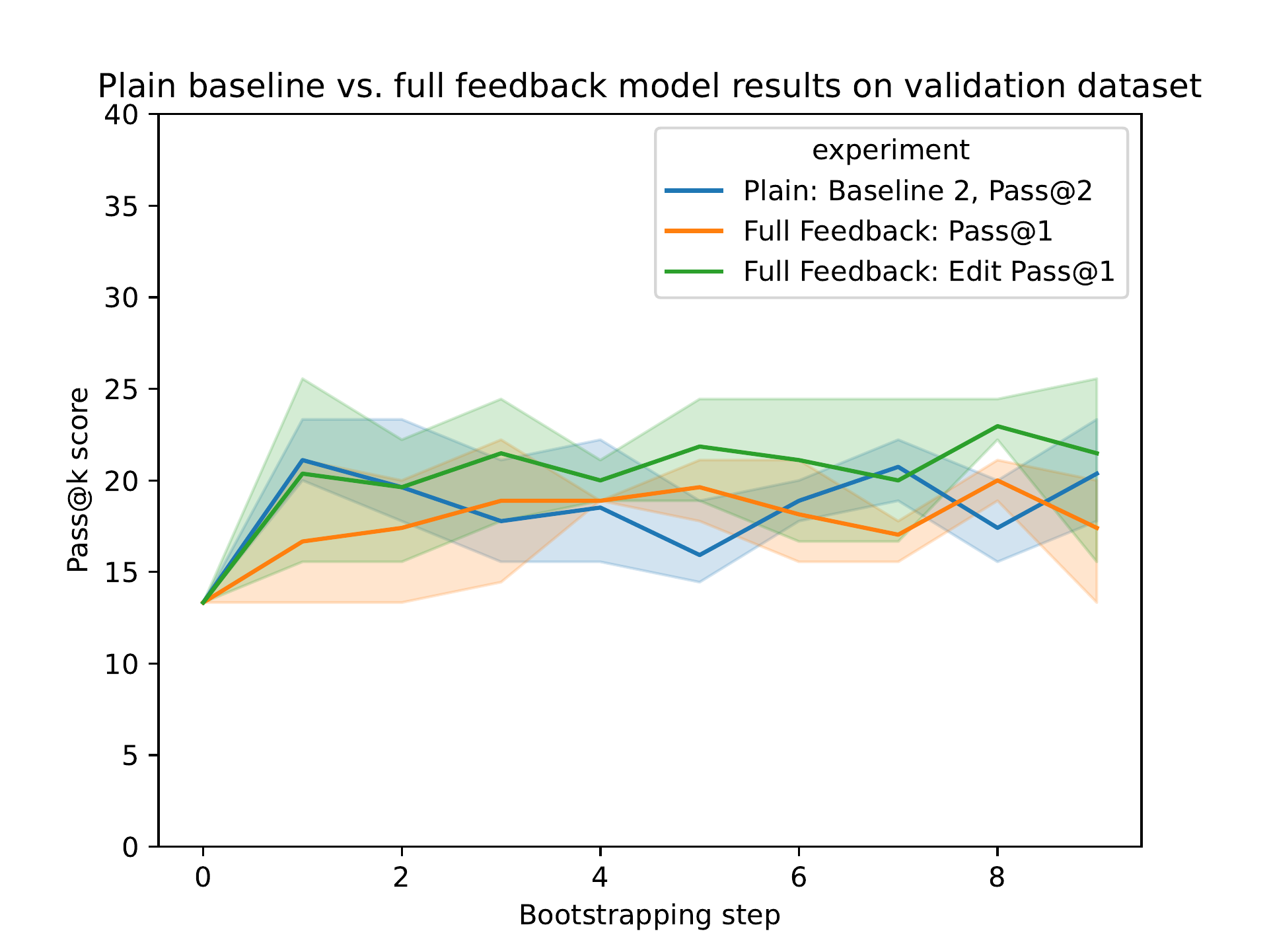}} \label{fig:mbpp_full_vs_baseline2}}%
    \caption{Results of Baseline 2 (beam search) compared to the repairing models while validating throughout the bootstrapping process. The samples come from 3 separate models trained with their respective training objective.}
    \label{fig:mbpp_validation_results_baseline2}%
\end{figure}

Selecting the best performing baseline of the plain bootstrapping model (Baseline 2), it can be compared to the repairing performance of the two repairing training objectives. This comparison throughout the bootstrapping process is shown in Figure \ref{fig:mbpp_validation_results_baseline2}.

Figure \ref{fig:mbpp_simple_vs_baseline2} and Figure \ref{fig:mbpp_full_vs_baseline2} show non-repairing and repairing performance of the simple and full feedback models respectively, compared to Baseline 2. Both repairing models seem to consistently outperform the baseline, with the full feedback model occasionally performing slightly worse than the baseline. \\

Overall, the bootstrapping training process does not show continuously increasing performance on the validation dataset for any training objective. For all training objectives there is large spike in pass@$k$ performance at bootstrapping step 1 (evaluation after the first fine-tuning step). Then there is a recession in performance for a few bootstrapping steps, only for performance to peak again later. 

\subsubsection{APPS}

\begin{figure}[H]
    \centering
    \subfloat[\centering Plain bootstrapping baselines.]{{\includegraphics[width=7.9cm]{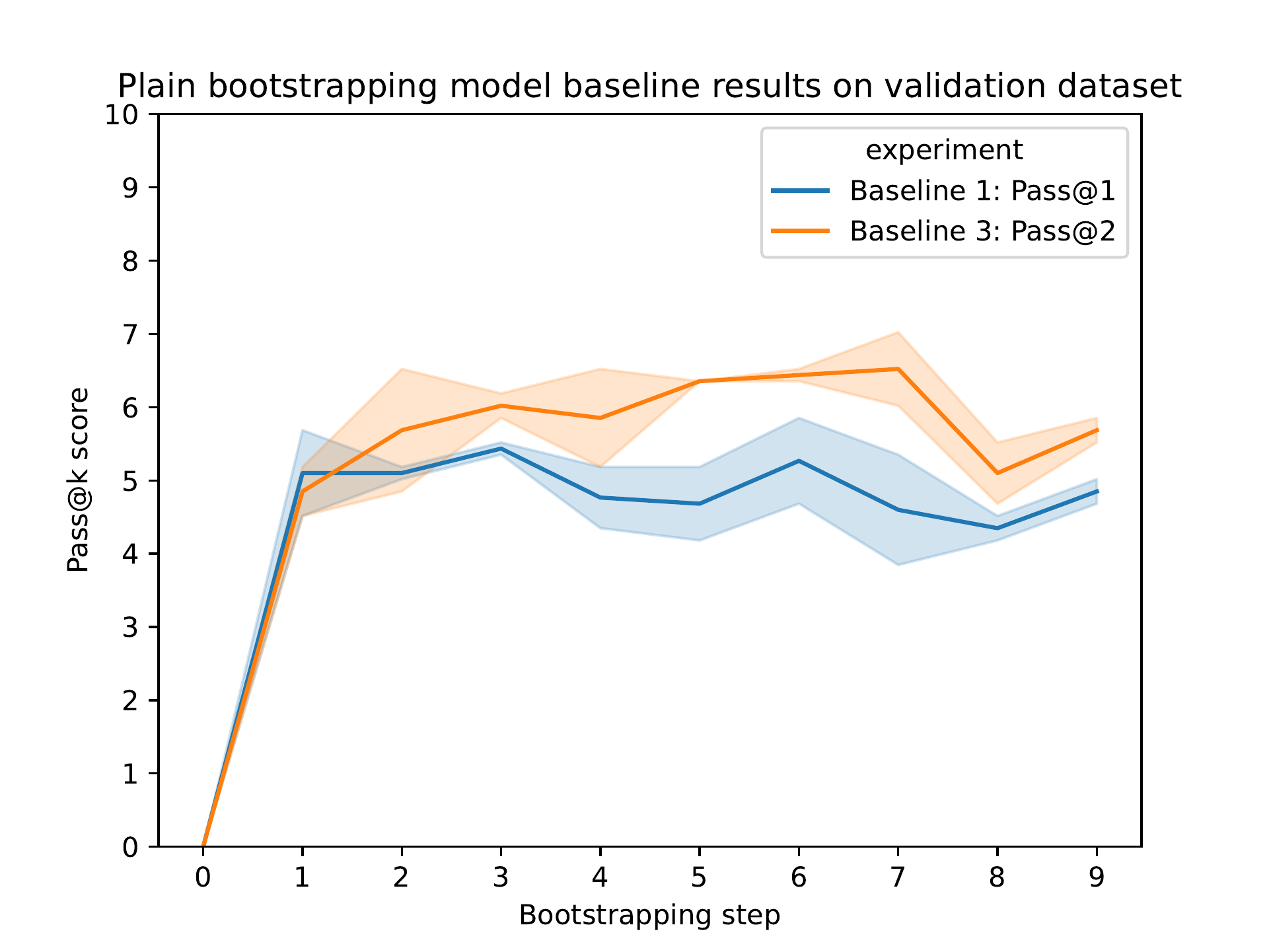} \label{apps_valid_plain_baselines} }}%
    \qquad
    \subfloat[\centering Baseline 1 of the plain bootstrapping model, compared to non-repairing performance of the full feedback repairing model.]{{\includegraphics[width=7.9cm]{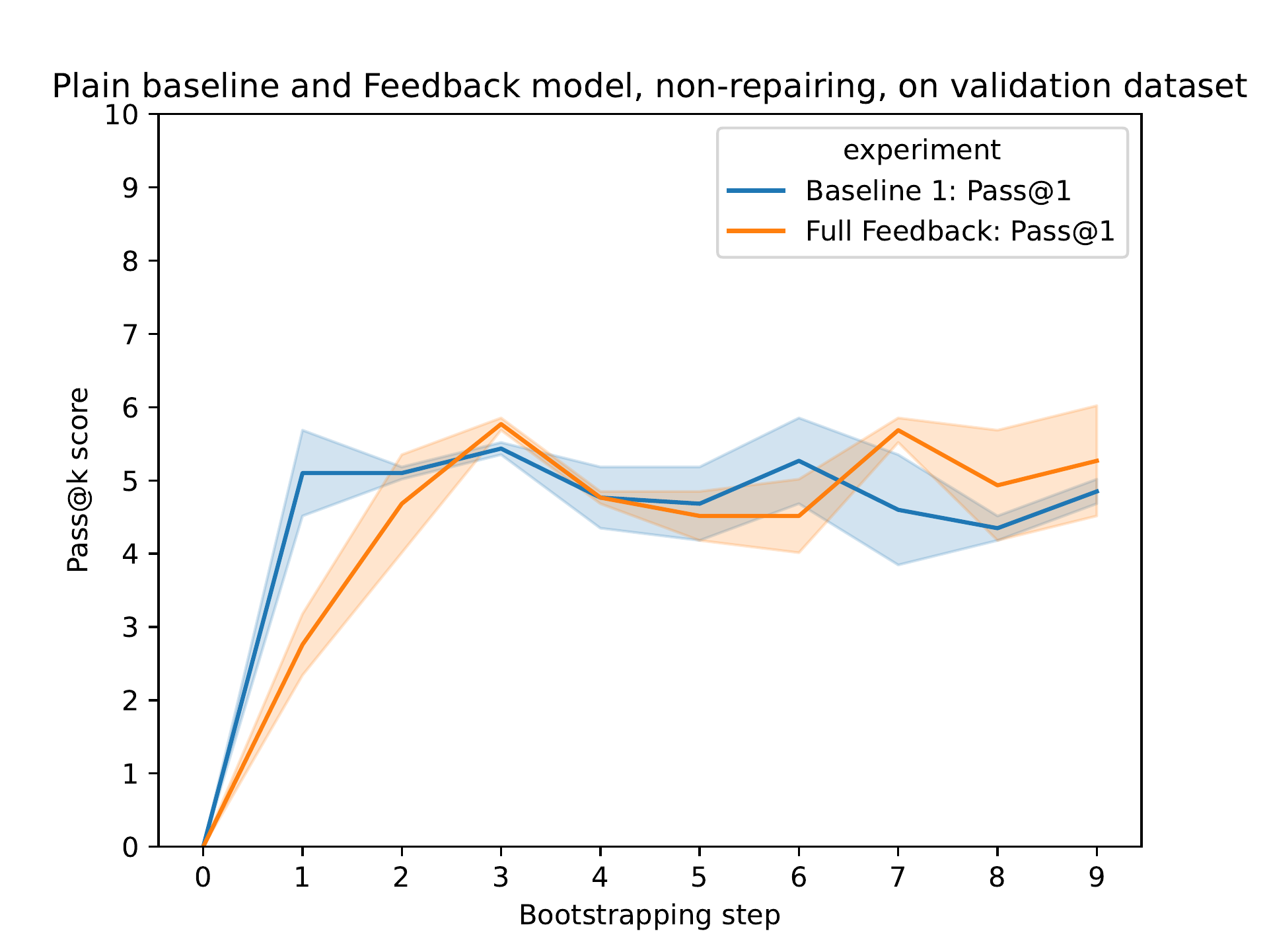}\label{apps_valid_baseline_1_vs_full_feedback_non_repairing} }}%
    \caption{Results of several experiments conducted while validating. The samples come from 2 separate models trained with their respective training objective.}
    \label{fig:apps_validation_results}%
\end{figure}

To compare baseline performance of the plain bootstrapping models throughout the bootstrapping process, Figure \ref{apps_valid_plain_baselines} has been plotted. Baseline 3 (temperature sampling, returning 2 solutions) performs best, and will therefore be used in comparisons with the repairing models. Notably, Baseline 3 seems to improve (almost) consistently until the 7th bootstrapping step, despite being more stochastic than Baseline 1. \\

Furthermore, Figure \ref{apps_valid_baseline_1_vs_full_feedback_non_repairing} shows non-repairing performance of the plain bootstrapping models compared to that of the full feedback repairing models. Importantly, unlike on the MBPP dataset, there appears to be a clear upward trend during the first 3 bootstrapping steps for the full feedback repairing model's initial pass@1 performance. The performance of the plain bootstrapping models quickly saturates, while the full feedback models keep improving for a few bootstrapping steps.

\begin{figure}[H]
    \centering
    \includegraphics[width=120mm]{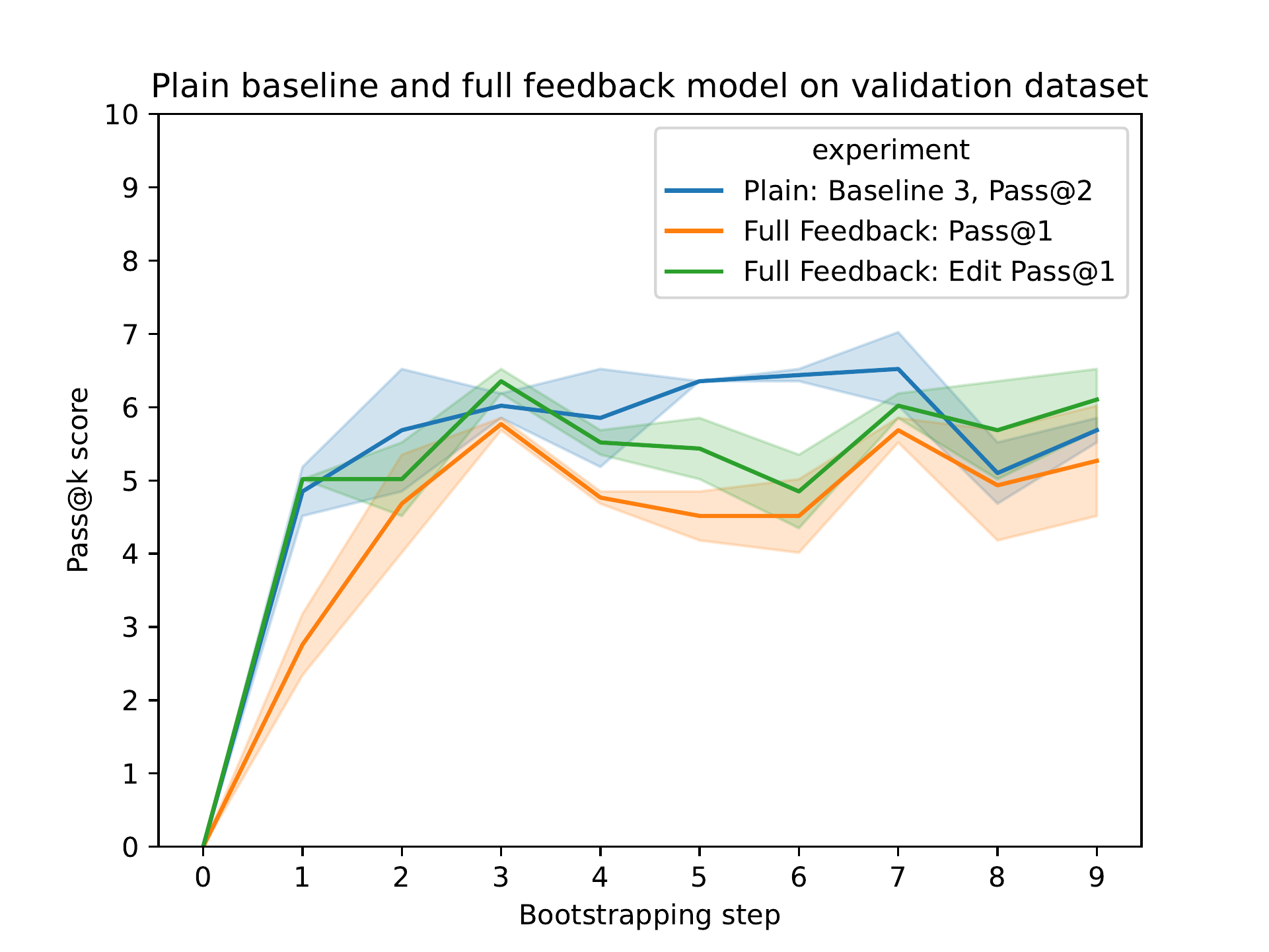}
    \caption{Baseline 3 of the plain model and pass@1 and repairing performance of the full feedback model.}
    \label{fig:apps_valid_baseline_3_vs_full_feedback}
\end{figure}

Figure \ref{fig:apps_valid_baseline_3_vs_full_feedback} uses the best performing baseline and compares it to the repairing performance of the full feedback model. Unlike on the MBPP dataset, the plain bootstrapping performs better, reaching a higher peak and improving more consistently.

\newpage

\subsection{Testing}
In this section, the testing performance of the best models (chosen from validation performance, see Appendix \ref{appendix:chosen_models}), is shown and analyzed. Both greedy performance and temperature sampled performance as in \citet{zhang_self-edit_2023} will be shown.

\subsubsection{MBPP}

\paragraph{Greedy performance} Table \ref{table:mbpp_greedy_performance} shows the greedy performance of all models with different training objectives. The baselines for models that used them are also displayed. To fairly compare the non-repairing models to the repairing models, one should compare Baseline 2 or Baseline 3 Pass@2 performance with Simple/Full Edit Pass@1 performance of the repair model that was trained with this prompt. For example, comparing the best performing baseline: Baseline 2 to the Simple Edit Pass@1 performance of the simple feedback model, there is a 3.0\% improvement when repairing while bootstrapping. \\

Solely looking at the greedy pass@1 performance between the regularly fine-tuned and plainly bootstrapped models, we observe a 5.7\% improvement over regular fine-tuning when having done plain bootstrapping. Comparing pass@1 performance between the plain bootstrapping models and full feedback models we observe an improvement of 16.3\% when having bootstrapped with repairing. This shows that even without repairing during inference, having trained on repairing improves performance by a large amount. \\

The regularly fine-tuned and plainly bootstrapped models perform poorly at repairing while not having been trained to repair, as evident from the edit performance in the second to last and last column. The two baselines seem to perform much better than greedily decoding these few-shot prompted repairing attempts. \\

Interestingly, the simple feedback models, using the full feedback few-shot prompt, are the best performing out of all. Prompted with these few-shot examples they in fact outperform the full feedback models that have been trained with this prompt. The simple feedback models appear to be better repairers in general, as evidenced by the difference between the pass@1 and edit pass@1 in both columns.

\begin{table}[H]
\resizebox{\columnwidth}{!}{
\begin{tabular}{l|lll|l
>{\columncolor[HTML]{F2F2F2}}l |l
>{\columncolor[HTML]{F2F2F2}}l }
\toprule
\textbf{Training Objective}   & \textbf{Pass@1} & \textbf{Baseline 2 Pass@2} & \textbf{Baseline 3 Pass@2} & \textbf{Simple Pass@1} & \textbf{Simple Edit Pass@1} & \textbf{Full Pass@1}  & \textbf{Full Edit Pass@1} \\ \toprule
Regular Fine-tuning           & 15.07 ± 0.96    & -                          & 16.80 ± 0.43               & -                      & -                           & 14.80 ± 0.57          & 14.87 ± 0.62              \\
Plain Bootstrapping           & 15.93 ± 0.62    & 19.73 ± 0.77               & 18.73 ± 0.81               & 15.00 ± 0.33           & 15.60 ± 0.00                & 14.67 ± 1.33          & 15.00 ± 1.18              \\
Simple Feedback Bootstrapping & -               & -                          & -                          & 17.47 ± 1.84  & 20.33 ± 2.39       & \textbf{18.53 ± 1.15} & \textbf{21.40 ± 2.12}     \\
Full Feedback Bootstrapping   & -               & -                          & -                          & 17.13 ± 0.25           & 18.67 ± 0.34                & 18.40 ± 0.28          & 19.93 ± 0.09 \\
\bottomrule
\end{tabular}
}
\caption{Greedy pass@1 and greedy edit pass@1 performance on the MBPP test set. Baseline 2 is beam search with a beam of 2 and Baseline 3 is temperature sampling with 2 samples (these results are sampled pass@2). The mean and standard deviation are reported based on 3 samples from 3 different models. The results in the first 3 columns use the ``Plain" few-example prompts. In the ``Simple" and ``Full" columns, the simple feedback and full feedback few-shot prompts have been used respectively. For the difference between repairing models, see Section \ref{method:trained_repairing_models}.}
\label{table:mbpp_greedy_performance}
\end{table}

\paragraph{Temperature sampling}
Additionally, Table \ref{table:mbpp_estimated_pass_at_k} shows the estimated pass@$k$ performance when doing temperature sampling. Here the performance gap between the regular fine-tuning and plain bootstrapping models disappears. The performance gap between the two repairing models also seems negligible, with the full feedback model occasionally performing better without repairing. \\

Comparing the plain bootstrapping models to the full feedback bootstrapping models on pass@10 performance, we observe a 7.2\% improvement using the full feedback models. When calculating the improvement of full feedback bootstrapping over plain bootstrapping when repairing (edit pass@10), we see a 21.6\% improvement. \\

\label{experiments:mbpp_results}To investigate the performance benefit of repairing, one should compare the Edit Pass@$k$ column with the regular Pass@$2k$ column. This is because the Edit Pass@$k$ column has had maximally twice as many tries at solving the problems than the Pass@$k$ column. Comparing the gray columns with the neighboring columns on the right, we observe that in almost all cases, simply generating twice the number of solutions outperforms repairing old solutions. However, having bootstrapped with repairing gives stronger performance overall, both for pass@$k$, edit pass@$k$, and increased repairing performance (performance from pass@$k$ to edit pass@$2k$).

\begin{table}[H]
\resizebox{\columnwidth}{!}{
\begin{tabular}{l|l
>{\columncolor[HTML]{F2F2F2}}l l
>{\columncolor[HTML]{F2F2F2}}l l
>{\columncolor[HTML]{F2F2F2}}l l
>{\columncolor[HTML]{F2F2F2}}l }
\toprule
Training Objective             & \textbf{Pass@1} & \textbf{Edit Pass@1} & \textbf{Pass@2} & \textbf{Edit Pass@2} & \textbf{Pass@5} & \textbf{Edit Pass@5} & \textbf{Pass@10} & \textbf{Edit Pass@10} \\ \toprule
Regular Fine-tuning (Plain Prompt) &  10.67 ± 0.58 & - & 16.77 ± 0.74 & - & 26.28 ± 0.97 & - &33.53 ± 1.20 & - \\
Regular Fine-tuning (Full Prompt)  & 10.05 ± 0.57 &  11.07 ± 0.54 & 15.73 ± 0.76 & 17.12 ± 0.71 & 24.92 ± 0.98 & 26.81 ± 0.87 & 32.13 ± 1.48 & 34.40 ± 1.23   \\ \midrule
Plain Bootstrapping (Plain Prompt) &  11.44 ± 0.18 & - & 17.25 ± 0.25 & - & 26.24 ± 0.41 & - & 33.27 ± 0.57 & - \\
Plain Bootstrapping (Full Prompt)  &  10.82 ± 0.23 &  11.85 ± 0.35 & 16.63 ± 0.22 & 18.02 ± 0.51 & 25.50 ± 0.11 & 27.29 ± 0.55 & 32.27 ± 0.09 & 34.33 ± 0.57 \\ \midrule
Simple Feedback Bootstrapping      & \textbf{13.50 ± 1.11} &  \textbf{18.80 ± 0.77} & \textbf{19.90 ± 1.12} & \textbf{25.99 ± 0.53} & 28.82 ± 0.83 & \textbf{35.34 ± 0.24} & 35.07 ± 0.41 & \textbf{41.73 ± 0.93} \\
Full Feedback Bootstrapping        &  13.26 ± 0.60 &  18.05 ± 0.68 & 19.74 ± 0.68 & 25.30 ± 0.78 & \textbf{28.97 ± 0.37} & 34.80 ± 0.63 & \textbf{35.67 ± 0.25} & 41.60 ± 0.43 \\
\bottomrule   
\end{tabular}
}
\caption{Estimated pass@$k$ results on the MBPP test dataset. These results have been generated using temperature sampling with $t=0.8$ and $n=10$. The mean and standard deviation are reported using the results of 3 different models per training objective.}
\label{table:mbpp_estimated_pass_at_k}
\end{table}

\begin{figure}[H]
    \centering
    \subfloat[\centering Full Feedback, Model: 1.]{{\includegraphics[width=4.6cm]{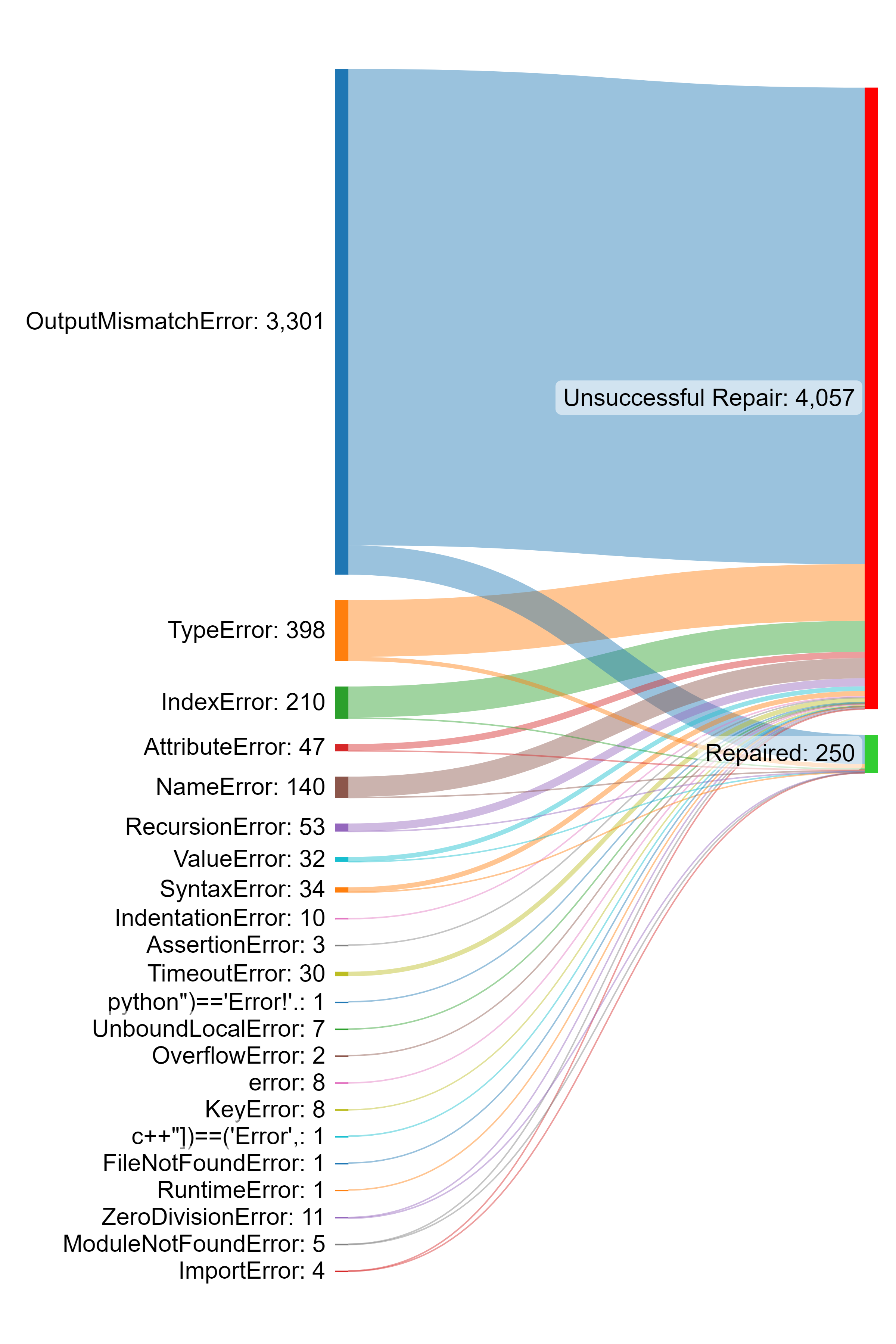} }}%
    \qquad
    \subfloat[\centering Full Feedback, Model: 2.]{{\includegraphics[width=4.6cm]{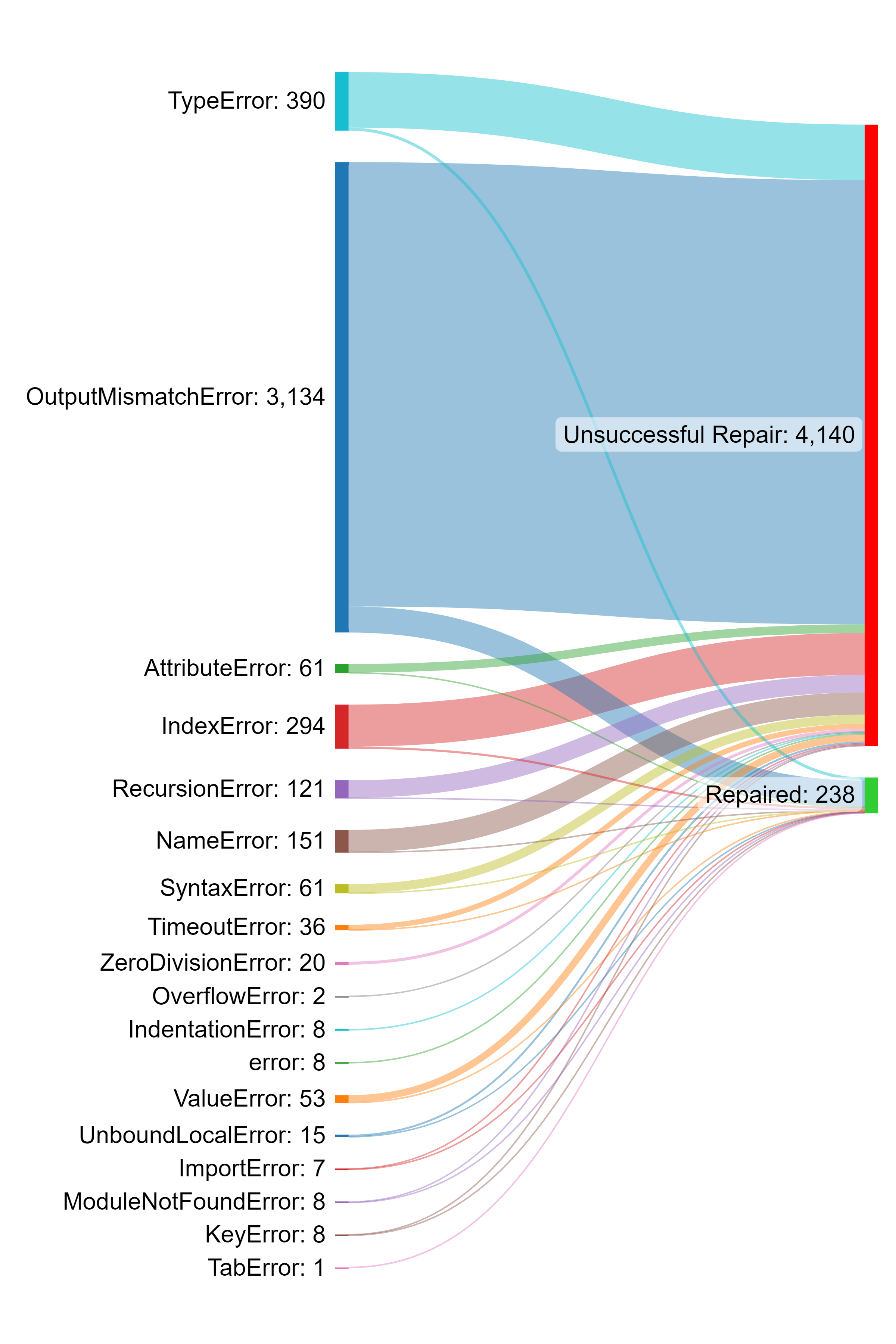} }}%
    \qquad
    \subfloat[\centering Full Feedback, Model: 3.]{{\includegraphics[width=4.6cm]{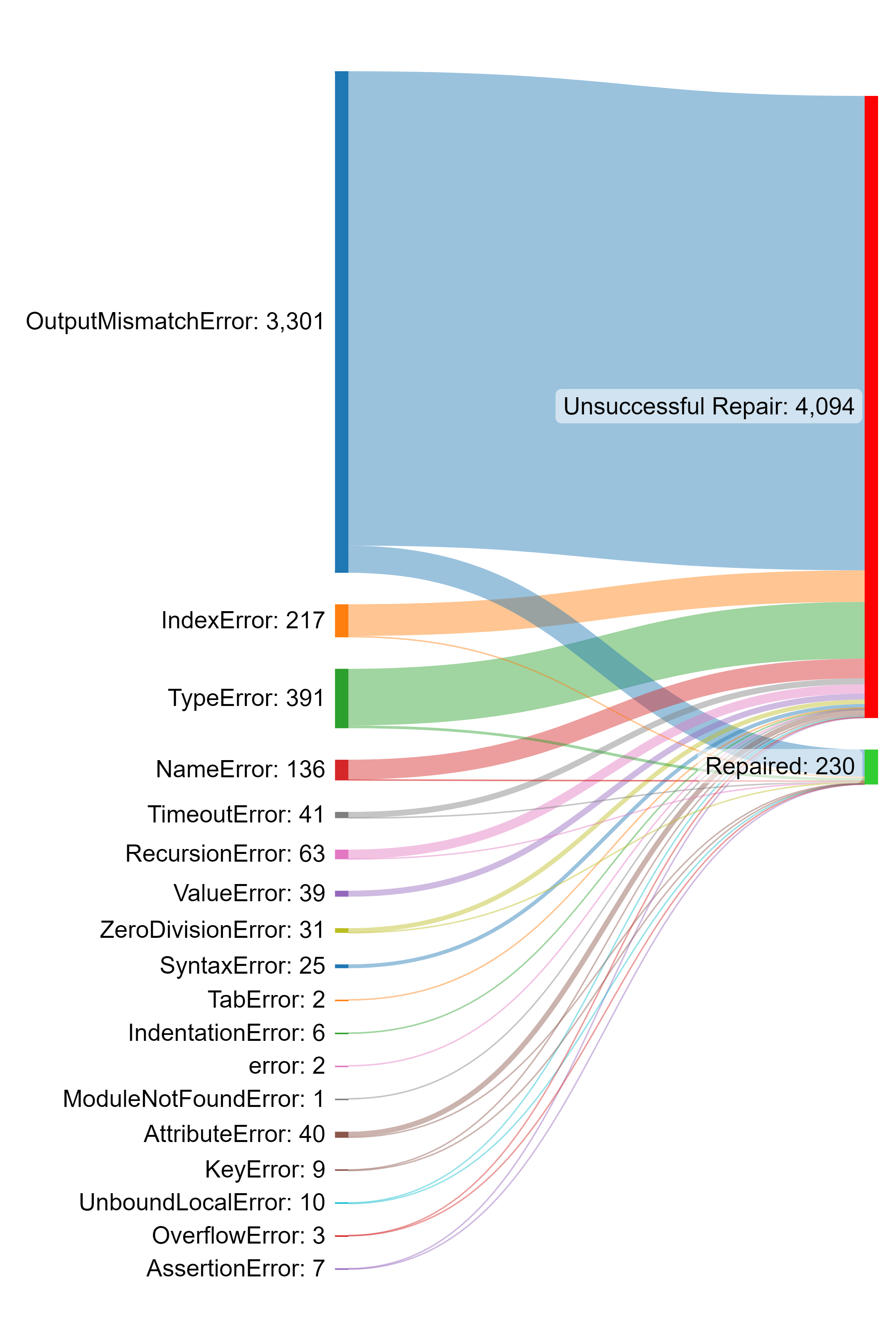} }}%
    \caption{Sankey diagrams of repairing performance on the MBPP test set while sampling, with $n=10$.}
    \label{fig:validation_sankey_diagram_repaired_errors}%
\end{figure}

\paragraph{Analyzing errors} Finally, in Figure \ref{fig:validation_sankey_diagram_repaired_errors} we also investigate the type of errors the repair models make and are able to repair. Since only the full feedback models receive feedback about error types (when it failed to produce correct code at the first synthesis attempt), we will only show plots for these models. The errors are standard Python errors, except for \texttt{OutputMismatchError} and those that are parsed incorrectly (``\texttt{python")=='Error!')}" and ``\texttt{c++")==('Error'}"). \texttt{OutputMismatchError} represents errors where the model is informed that the output produced by the synthesized function does not match the expected output (on the unit tests). The model is also informed what the produced output was, unlike on APPS.

Notably, the 3 models appear to produce approximately the same errors, with in all cases \texttt{OutputMismatchError}, \texttt{TypeError}, \texttt{IndexError}, and \texttt{NameError} being the most frequent errors, in that order. The repairing attempts of the different errors also succeeds in approximately the same manner. \texttt{OutputMismatchError} leads to the largest portion of repaired codes, while also being the most frequent error by a large margin.

\subsubsection{APPS}

\paragraph{Greedy performance} Table \ref{table:apps_greedy_performance} shows the greedy performance of all models with different training objectives on the APPS dataset. The plain bootstrapping models outperform the models with different training objectives. Notably, the full feedback models perform poorly compared to the other models. However, the improvement from repairing (edit pass@1 performance improvement over pass@1) is 63.4\%. Additionally, plain bootstrapping performs 17.5\% better than doing regular fine-tuning at pass@1.

\begin{table}[H]
\begin{tabular}{l|l >{\columncolor[HTML]{F2F2F2}}l}
\toprule
\textbf{Training Objective} & \textbf{Pass@1} & \textbf{Edit Pass@1} \\
\toprule
Regular Fine-tuning         & 0.63 ± 0.08     & 0.66 ± 0.09       \\
Plain Bootstrapping         & \textbf{0.74 ± 0.03}     & \textbf{0.79 ± 0.04 }      \\
Full Feedback Bootstrapping & 0.41 ± 0.06     & 0.67 ± 0.12       \\
\bottomrule
\end{tabular}
\caption{Greedy pass@1 and greedy edit pass@1 results of the trained models on the test dataset. The mean and standard deviation are reported based on 2 samples from 2 different models.}
\label{table:apps_greedy_performance}
\end{table}

\paragraph{Temperature sampling} Table \ref{table:apps_estimated_pass_at_k} shows the estimated pass@$k$ and edit pass@$k$ performance. The edit pass@$k$ performance of the repairing model is almost equivalent to simply sampling twice the number of solutions (unlike in the MBPP results). The plain bootstrapping model shows superior performance here too. The full feedback repairing model still performs poorly compared to the other two. However, on repair attempts the full feedback model outperforms the regular fine-tuning model.

\begin{table}[H]
\resizebox{\columnwidth}{!}{
\begin{tabular}{l|l
>{\columncolor[HTML]{F2F2F2}}l l
>{\columncolor[HTML]{F2F2F2}}l l
>{\columncolor[HTML]{F2F2F2}}l l
>{\columncolor[HTML]{F2F2F2}}l}
\cline{2-9}
\toprule
\textbf{Training Objective} & \textbf{Pass@1} & \textbf{Edit Pass@1} & \textbf{Pass@2} & \textbf{Edit Pass@2} & \textbf{Pass@5} & \textbf{Edit Pass@5} & \textbf{Pass@10} & \textbf{Edit Pass@10} \\ \toprule
Regular Fine-tuning         &  0.22 ± 0.08 & 0.31 ± 0.09 & 0.41 ± 0.15 & 0.55 ± 0.16 & 0.83 ± 0.28 & 1.09 ± 0.29 & 1.32 ± 0.38 & 1.68 ± 0.41 \\
Plain Bootstrapping         & \textbf{0.28 ± 0.04} & \textbf{0.39 ± 0.03} & \textbf{0.51 ± 0.06} & \textbf{0.70 ± 0.05} & \textbf{1.01 ± 0.10} & \textbf{1.36 ± 0.07} & \textbf{1.57 ± 0.07} & \textbf{2.10 ± 0.01}             \\
Full Feedback Bootstrapping &  0.19 ± 0.02          & 0.34 ± 0.06          & 0.35 ± 0.04          & 0.59 ± 0.10          & 0.71 ± 0.08          & 1.14 ± 0.19          & 1.13 ± 0.14          & 1.76 ± 0.30 \\
\bottomrule
\end{tabular}
}
\caption{Estimated pass@$k$ and edit pass@$k$ results on the APPS test dataset. These results have been generated using temperature sampling with $t=0.8$. The mean and standard deviation are reported using the results of 2 different models per training objective.}
\label{table:apps_estimated_pass_at_k}
\end{table}

\paragraph{Comparison to \citet{zhang_self-edit_2023}} Since the results by \citet{zhang_self-edit_2023} work with sampled pass@$k$, Table \ref{table:apps_big_comparison} compares the sampled pass@$k$ and edit pass@$k$ performance with the models tested by \citet{zhang_self-edit_2023}. Our results between estimated and sampled pass@$k$ do not seem to differ substantially. The major findings in the previous table seem to hold up here too. One key difference between the models of \citet{zhang_self-edit_2023} and ours is that they make use of all ground-truth solutions during training, rather than only the first like we do. Besides, they do not remove the task cases from the training and validation prompts.

\begin{table}[H]
\resizebox{\columnwidth}{!}{
\begin{tabular}{l|l
>{\columncolor[HTML]{F2F2F2}}l l
>{\columncolor[HTML]{F2F2F2}}l l
>{\columncolor[HTML]{F2F2F2}}l l
>{\columncolor[HTML]{F2F2F2}}l }
\toprule
Code Gen. Model             & \textbf{Pass@1} & \textbf{Edit Pass@1} & \textbf{Pass@2} & \textbf{Edit Pass@2} & \textbf{Pass@5} & \textbf{Edit Pass@5} & \textbf{Pass@10} & \textbf{Edit Pass@10} \\ \toprule
\textbf{Fine-tuned (\citeauthor{zhang_self-edit_2023})}          &                 &                      &                 &                      &                 &                      &                  &                       \\
PyCodeGPT (110M)  & 0.20 & 0.64 & - & - & 0.38 & 0.98 & 0.44 & 1.24                       \\
GPT-Neo 125M                &  0.08 & 0.22 & - & - & 0.40 & 0.70 & 0.70 & 1.12                       \\
CodeGen 350M                &  0.20 & 0.32 & - & - & 0.30 & 0.56 & 0.32 & 0.84 \\
GPT-Neo 1.3B                &  0.14 & 0.68 & - & - & 0.74 & 1.38 & 1.40 & 2.10 \\
InCoder 1.3B                &  0.66 & 0.86 & - & - & 1.18 & 1.62 & 1.44 & 2.10   \\
GPT-J (6B) &\textbf{0.70} & \textbf{1.40} & - & - & \textbf{2.46} & \textbf{3.34} & \textbf{3.52} & \textbf{4.76} \\ \midrule
\textbf{Zero-shot (\citeauthor{zhang_self-edit_2023})} & \textbf{}       & \textbf{}            & \textbf{}       & \textbf{}            & \textbf{}       & \textbf{}            & \textbf{}        & \textbf{}             \\ 
InCoder 1.3B & 0.00 & 0.24 & - & - & 0.02 & 0.50 & 0.02 & 0.76 \\
CodeGen 2.7B & 0.12 & 0.28 & - & - & 0.34 & 0.66 & 0.66 & 1.08 \\
GPT3-davinci-002 (175B) & \textbf{7.48} & \textbf{7.94} & - & - & \textbf{15.94} & \textbf{16.66} & - & - \\ 
\midrule   
\textbf{Fine-tuned CodeT5 (770M)} & \textbf{}       & \textbf{}            & \textbf{}       & \textbf{}            & \textbf{}       & \textbf{}            & \textbf{}        & \textbf{}             \\

\multicolumn{1}{r|}{Regular Fine-tuning}       & 0.14 ± 0.08     & 0.22 ± 0.14          & 0.34 ± 0.16     & 0.48 ± 0.21          & 0.87 ± 0.30     & 1.08 ± 0.27          & 1.32 ± 0.38      & 1.68 ± 0.41           \\
\multicolumn{1}{r|}{Plain Bootstrapping}         & \textbf{0.33 ± 0.07}     & \textbf{0.44 ± 0.09 }         & \textbf{0.56 ± 0.11}     & \textbf{0.72 ± 0.11 }         & \textbf{1.05 ± 0.12}     & \textbf{1.44 ± 0.12}          & \textbf{1.57 ± 0.07}     & \textbf{2.10 ± 0.01}           \\
\multicolumn{1}{r|}{Full Feedback Bootstrapping} & 0.18 ± 0.00     & 0.38 ± 0.08          & 0.26 ± 0.04     & 0.61 ± 0.20          & 0.65 ± 0.06     & 1.10 ± 0.21          & 1.13 ± 0.14      & 1.76 ± 0.30           \\
\bottomrule   
\end{tabular}
}
\caption{Sampled pass@$k$ results on the APPS test dataset. The fine-tuned CodeT5 (770M) results are the baselines and repairing training objective proposed by our work. These results have been generated using temperature sampling with $t=0.8$. The mean and standard deviation are reported using the results of 2 different models per training objective.}
\label{table:apps_big_comparison}
\end{table}

Notably, our plain bootstrapping models outperform models of much larger sizes (in parameter size) such as: GPT-Neo 1.3B and InCoder 1.3B (Pass@10), which both have 68.8\% more parameters. While the regular-finetuning models lack behind (as well as the repairing model). Initially, the pass@$k$ performance of InCoder 1.3B is twice as good as that of the plain bootstrapping models when k is small. However, this lead diminishes as more samples are drawn.  \\

Comparing edit pass@5 and pass@10 of the models fine-tuned by \citet{zhang_self-edit_2023}, we observe that for both GPT-Neo models and GPT-J, repairing does not actually improve performance more than simply sampling twice as much. Hence, this is not a finding unique to our method.

\paragraph{Comparison to \citet{le_coderl_2022}}\label{results:coderl_performance} In Table \ref{table:apps_coderl_comparison} we compare our results with the estimated pass@$k$ results of the CodeRL paper \cite{le_coderl_2022}. The fine-tuned CodeT5 model by \citet{le_coderl_2022} without CodeRL already severely outperforms our models and even GPT-J (a 6B parameter model) fine-tuned by \citet{zhang_self-edit_2023}. CodeRL performs even better, leading to extremely high pass@$k$ scores. The possible reasons for this discrepancy between scores will follow in the discussion Section \ref{discussion:codeRl_performance_discrepancies}.

\begin{table}[H]
\resizebox{\columnwidth}{!}{
\begin{tabular}{l|l
>{\columncolor[HTML]{F2F2F2}}l l
>{\columncolor[HTML]{F2F2F2}}l l
>{\columncolor[HTML]{F2F2F2}}l l
>{\columncolor[HTML]{F2F2F2}}l }
\toprule
Code Gen. Model             & \textbf{Pass@1} & \textbf{Edit Pass@1} & \textbf{Pass@2} & \textbf{Edit Pass@2} & \textbf{Pass@5} & \textbf{Edit Pass@5} & \textbf{Pass@10} & \textbf{Edit Pass@10} \\ \toprule
\textbf{CodeRL (\citeauthor{le_coderl_2022})} & \textbf{}       & \textbf{}            & \textbf{}       & \textbf{}            & \textbf{}       & \textbf{}            & \textbf{}        & \textbf{}             \\
Fine-tuned CodeT5 (770M) & 2.00       & -      & -       & -            & 2.90      & - &  -        & -    \\
CodeRL+CodeT5 (?B) & \textbf{2.69}       & -      & -       & -            & \textbf{6.81}      & - &  -        & -             \\

\midrule   
\textbf{Fine-tuned CodeT5 (770M)} & \textbf{}       & \textbf{}            & \textbf{}       & \textbf{}            & \textbf{}       & \textbf{}            & \textbf{}        & \textbf{}             \\

\multicolumn{1}{r|}{Regular Fine-tuning} &  0.22 ± 0.08 & 0.31 ± 0.09 & 0.41 ± 0.15 & 0.55 ± 0.16 & 0.83 ± 0.28 & 1.09 ± 0.29 & 1.32 ± 0.38 & 1.68 ± 0.41 \\
\multicolumn{1}{r|}{Plain Bootstrapping}         & \textbf{0.28 ± 0.04} & \textbf{0.39 ± 0.03} & \textbf{0.51 ± 0.06} & \textbf{0.70 ± 0.05} & \textbf{1.01 ± 0.10} & \textbf{1.36 ± 0.07} & \textbf{1.57 ± 0.07} & \textbf{2.10 ± 0.01} \\
\multicolumn{1}{r|}{Full Feedback Bootstrapping} & 0.19 ± 0.02          & 0.34 ± 0.06          & 0.35 ± 0.04          & 0.59 ± 0.10          & 0.71 ± 0.08          & 1.14 ± 0.19          & 1.13 ± 0.14          & 1.76 ± 0.30 \\
\bottomrule   
\end{tabular}
}
\caption{Comparison of CodeRL's estimated pass@$k$ with our estimated pass@$k$ performance on the APPS test dataset.}
\label{table:apps_coderl_comparison}
\end{table}

\newpage

\section{Discussion}

In this section, we will discuss some of the most important results to answer the research question posed in the introduction. Limitations of the methodology and future work will also be discussed.

\subsection{Few-shot prompting influencing pre-trained model performance on MBPP}
In the validation results section on the MBPP dataset \ref{results:prompting_influences_pre_trained_model}, we noted that prompting the pre-trained model with different few-shot prompts impacts performance. \\

The full feedback prompt gives the best performance. Attempting to understand what causes this difference in performance is not trivial. One might argue that having more code examples in the prompt (as compared to the plain bootstrapping prompts) could be the cause. However, the pre-trained model performs worst with the simple feedback prompts (which has the same number of code examples as that of the full feedback prompts). Another theory is that the model benefits from seeing the output of the few-shot examples. It is also possible that these performance differences are purely coincidental. A way to investigate this quantitatively would be to produce more of these prompts with different coding solutions (but using the same solutions per training-objective prompt) and validating with them. \\

Nonetheless, since the plain bootstrapping prompt performs slightly better than the simple feedback prompt, we do not expect the prompts to favor the repairing models' performance. However, it is important to keep in mind that the repairing prompts do contain more information.

\subsection{Performance gains of bootstrapping and bootstrapping with repairing}
The modest improvement of the plain bootstrapping results on the MBPP dataset over regular fine-tuning (greedy Pass@1, shown in Table \ref{table:mbpp_greedy_performance}) suggests there is merit to the bootstrapping method (at least on MBPP, which is a small dataset). \\

In turn, the larger improvement of the repairing models over the plain bootstrapping models (greedy Pass@1 compared to greedy Full Pass@1, Table \ref{table:mbpp_greedy_performance}), shows there is merit to repairing while bootstrapping (on MBPP). \\

Conversely, on the APPS dataset bootstrapping while repairing seems to give worse performance than simply plain bootstrapping (as shown in Table \ref{table:apps_greedy_performance}). However, on this dataset plain bootstrapping performs much better than regular fine-tuning. Hence, on both datasets plain bootstrapping outperforms regular-finetuning, presenting bootstrapping as a way to improve programming performance. \\

The benefit of repairing while bootstrapping (on pass@1 performance) is thus unclear. There are many factors that might have affected this discrepancy of programming performance between the MBPP and APPS models. The MBPP dataset is much smaller (less data points), and generally seen as simpler (shorter tasks, shorter ground truth code). Additionally, the APPS dataset has been limited by cutting out the example tests during training (due to aforementioned problems). Successfully training these repair models (to perform well even without repairing) might depend heavily on being able to see the example test cases during training. 

It is also possible that, because the ground-truth solutions are only reinforced at the end of repairing, the original task is already outside of the context window on the APPS dataset. This is very likely to affect first pass performance. \\

Comparing to the related work, specifically the results given by \citet{zhang_self-edit_2023} and \citet{le_coderl_2022}, it is difficult to conclude which method is definitively better. The plain bootstrapping models outperform larger models trained by \citet{zhang_self-edit_2023}, suggesting it is a good method to outperform larger models. However, the same pre-trained model that we use, simply fine-tuned by \citet{le_coderl_2022} and using their reported results, outperforms all our training objectives by a large margin (more on this in \ref{discussion:codeRl_performance_discrepancies}). \\

Furthermore, no comparisons have been drawn to results achieved on MBPP in the literature. This is because usually MBPP is not used to train, but rather only as a validation dataset after training on a larger dataset like APPS. Besides, \citet{zhang_self-edit_2023} does not report on MBPP and \citet{le_coderl_2022} only reports on pass@80 and pass@1000, making it computationally very costly to compare. This further complicates conclusions about the actual performance gains from this method compared to other work. 

\subsection{Benefit of repairing during inference}

On the MBPP dataset, repairing rarely outperforms regularly sampling twice as much (this results in the same number of times the model is maximally sampled from), even on the repairing models, as evidenced by Table \ref{table:mbpp_estimated_pass_at_k}. \\

On APPS, repairing during inference comes close to, and sometimes surpasses, simply sampling twice as much (on the full feedback repairing models, see Table \ref{table:apps_estimated_pass_at_k}). However, the pass@1 performance of the full feedback models is weak compared to that of the other training objectives. \\

If repairing during inference does not seem to give a big improvement, which seems to be the case here, it would likely be faster and therefore preferred to regularly sample twice as much. This is because sampling $2k$ programs can be done in parallel, while first sampling $k$ programs (in parallel) and then feeding these programs to be repaired to the model and sampling $k$ programs again (in parallel) is a sequential operation, and therefore slower. However, this does not imply it should be discouraged during training as during training it can improve inference performance. \\

However, there could be one situation in which repairing during inference would be preferred when the edit pass@$k$ performance does not outperform the regular sampling pass@$2k$. This would be when the repair model already solved the problem with one of the $k$ problems that were initially sampled (judged on the unit tests). Then if we decide not to repair, we will have saved energy and perhaps even time (depending on GPU/TPU resources). Though, this sampling approach would not guarantee that the candidate program passes the hidden tests. Therefore, if the computational resources are present and energy is abundant, simply sampling models in parallel would still be preferred. \\

Having a smaller model do only the repairing as in \citet{zhang_self-edit_2023} is also an elegant way to save computational resources. However, this would complicate the bootstrapping process and potentially weaken the performance gains observed.

\subsection{CodeRL performance discrepancies}
\label{discussion:codeRl_performance_discrepancies}
As alluded to in Section \ref{results:coderl_performance} there seems to be a large gap in performance between the reported performance of the CodeRL models by \citet{le_coderl_2022} and those of \citet{zhang_self-edit_2023} and ours. 

Most surprising is that when regularly fine-tuning the same pre-trained model we get an estimated 0.22 pass@1 performance, while this results in an estimated 2.00 pass@1 performance in the CodeRL paper \cite{le_coderl_2022}. This performance gap is puzzling, as one would expect the results to come much closer, even when using slightly different hyperparameters. We will list the differences in methodology to try and understand what might have caused this difference. \\

First, the CodeRL models have been fine-tuned on the full dataset, training on all ground-truth examples and not splitting the dataset up in a validation dataset for early stopping \cite{le_coderl_2022}. 

Second, through correspondence with the main author of the CodeRL paper, we find that during inference they use a temperature of 0.6 with nucleus sampling $p=0.95$. They also sample $n=1000$ solutions, rather than $n=10$, for the pass@$k$ estimates. It is likely that sampling 100 times more solutions with a lower temperature and smarter sampling method will result in a higher pass@$k$ estimate. The fact that other models fine-tuned models tested by both papers differ substantially in performance seems to attest to this point (GPT-Neo: 0.14/1.12, GPT-J: 0.70/1.82 pass@1). However, this is unattainable with the limited resources available to us. The reason that $t=0.8$ and $n=10$ was chosen by us is because it matches the inference methodology of \citet{zhang_self-edit_2023} and was actually computationally feasible. \\

It is also possible that their hyperparameters/learning rate scheduling is vastly superior to ours and that ours are potentially poorly chosen. It could be that we left a large amount of the pre-trained model's potential untapped because of this. Copying their exact hyperparameters and learning rate scheduling could take at least this uncertainty away. 

\subsection{Limitations}
The limitations of this work have been discussed throughout the discussion, but they will be summarized and elaborated on here. \\

The largest limitation of bootstrapping in this way, fine-tuning on the pre-trained model every bootstrapping step, rather than improving the fine-tuned model, is the substantial computational cost. This makes the method in its current form almost infeasible for improving very large language models for program synthesis. 

One also has to be very careful with example test cases as the bootstrapping method will make models exploit them if they overlap with hidden tests. Sadly, this has happened multiple times during the development of this method. \\

Sampling 10 solutions to compare to other methods in the literature is likely not sufficient to give definitive answers to which method is best. There were also not many papers to directly compare to as repairing with large language models of code is still a very active and new field. It is also difficult to compare apples to apples when sizes of language models differ substantially. 

Similarly, hyperparameters could have been optimized for each training objective, instead of using the same hyperparameters that seemed to work well while fine-tuning and evaluating on the validation dataset for all of the training objectives. This might change the entire conclusion and is perhaps why the CodeRL performance is so high. However, using different hyperparameters for all the training objectives would have introduced another variable making a fair comparison harder. \\

Bootstrapping as a method could simply be making use of random variations in the training initialization. When doing 10 bootstrapping steps, you are fine-tuning the same pre-trained model 10 times (with different data). With the randomization of parameter initialization at the start of training, it could be the case that while bootstrapping you simply get more chances at throwing the die, which means more chances to get lucky with parameter initialization, leading to higher performance. To test this theory, one could also fine-tune 10 regular fine-tuning models and pick the best one (based on the validation set), or fix the training seed (as was attempted before but we did not succeed in). However, since the variance of the regular fine-tuning models does not appear extremely large compared to the other training objectives (Table \ref{table:mbpp_greedy_performance} and Table \ref{table:apps_greedy_performance}), it might not be too concerning. \\

Using a model with such a limited context window has likely affected the first pass performance of the repairing model on APPS substantially. The results would likely be more interesting with models of a larger context size, though these are usually more computationally expensive to run and train. \\

Finally, repairing seems to be inferior to simply sampling the same amount. However, repairing during bootstrapping training, at least on MBPP, seems to improve non-repairing performance over regular fine-tuning and plain bootstrapping. So there might still be benefits to using such a methodology during training.

\subsection{Research question}
In this thesis, we set out to answer the following research question: \textbf{how can self-taught programming and repairing be used to improve program synthesis performance of language models?}\\

Given the results of this thesis, it can be stated that we have introduced a bootstrapping algorithm (self-teaching) that outperforms regular fine-tuning on programming competition datasets (Algorithm \ref{algo:bootstrapping}). We have also shown how this algorithm can be adapted to support bootstrapping with repairing, to teach a language model how to repair. On the MBPP dataset, bootstrapping with repairing makes the models perform even better than regular bootstrapping, while on APPS it makes it worse. 

Finally, we showed how repairing during inference with our models might not be optimal. Even while training with repairing improves performance over other methods (like in MBPP), during inference we showed how simply sampling in parallel with our models can reach the same level of performance as repairing or better.

\newpage

\subsection{Future work}
Since program repair using large language models is an exciting and rapidly evolving field, we feel like there is no shortage of future work to conduct.\\

Most importantly, the testing methodology could be improved to increase the significance of our results. First, regarding regular fine-tuning, the pre-trained model should be fine-tuned as many times as the bootstrapped models and select the best models based on the validation performance. This should take away worries about different random initializations and the large difference in compute.
Second, more solutions for the pass@$k$ estimate should be generated, this would make the estimate much more accurate. Related to this, we should investigate how we can reach the same level of performance with regular fine-tuning as the CodeRL paper. Once these values are aligned, comparing the (repairing) bootstrapping results will tell us more about the benefit of the method discussed in this thesis, as we are using the same base model and datasets. \\ 

Additionally, there are several ways to improve the programming performance of the repair models. First, while doing repair bootstrapping, instead of only reinforcing the repaired or ground-truth example on the repairing prompt, we could also reinforce the solution on the first go prompt. This will likely improve the non-edit pass@$k$ performance of the repair models on APPS due to the short trained context length of CodeT5. Since the tasks in APPS are so long, while repairing the model will likely not see the full tasks, currently only reinforcing ground-truth answers on left-truncated prompts containing primarily incorrect code. This likely makes it very difficult for the model to produce correct code in one go. Second, hyperparameter optimizations could be conducted to optimize the performance of each training objective separately. Third, since different prompts seemed to affect the pre-trained models' performance on MBPP, we could search for better prompts (perhaps using \cite{xu_reprompting_2023}). \\

To improve the overall performance of the models one could consider using a pre-trained model that was trained with a larger context width. However, the bootstrapping with repairing algorithm might need to be adapted to support decoder-only models, as otherwise the model might produce faulty code on purpose otherwise.  \\

Moreover, to make the bootstrapping method, proposed in this thesis, practical for larger models, it will have to be optimized for such a use case. This is because there is an enormous computational cost to fine-tuning the pre-trained model every bootstrapping step. Which would become worse once the model is bigger. A logical first approach would be to fine-tune the previously fine-tuned model, making sure the early stopping does not trigger before at least a few updating steps. One could also experiment with using a replay-buffer, this prevents wasting previously generated data every iteration, which might help the bootstrapping process to converge more quickly. \\

Finally, investigating the repeated repairing performance of these models could also be an interesting experiment. When the model fails to repair, we can replace the first attempt with this last failed attempt and repeatedly perform this action until the model repairs correctly. This way we can see if multiple repairing attempts outperforms simply sampling more.

\chapter{Conclusion}
This work focused on two pressing matters within the field of program synthesis and attempted to solve them. The first problem was that programming competition datasets, used to fine-tune and evaluate pre-trained language models on, are limited in size and quality, while these language models are extremely data hungry. Since human annotation is labor intensive and requires expertise, we would benefit from models teaching themselves. The second issue was that the program synthesis language models have a misaligned program synthesis process as compared to humans. While humans iteratively develop code with the help of a compiler, most program synthesis models currently produce code in one go. \\

In order to overcome the challenge of data scarcity we proposed a bootstrapping method for program synthesis models. Additionally, to increase alignment of the program synthesis process, we show how this algorithm can be used to teach models how to repair their own output during the bootstrapping process. \\

We investigated the performance of several baselines and our bootstrapping and repairing methods on the MBPP and APPS dataset. This showed that bootstrapping improves pass@$k$ performance over regular-finetuning in all cases. We also show that training with repairing while bootstrapping improves non-repairing performance on the non-faulty MBPP dataset over plain bootstrapping substantially. Additionally, we conclude that repairing during inference might not be advantageous for our models. We found that regularly sampling twice the number of initial solutions and not repairing gives the same or better performance in most cases (while faster). \\ 

Furthermore, we have shown several problems with the APPS dataset relating to the unit tests. The unit tests overlap in the train dataset, making bootstrapping prone to overfitting when they are not removed. Some unit tests also seem to fail without reason on the code validation script provided by \citet{le_coderl_2022}, even on ground-truth solutions, furthermore confusing the bootstrapping process if they are not removed.
Additionally, we have pointed out that most of the task descriptions far exceed the trained context length of the CodeT5 model.

These significant findings about the APPS dataset will be valuable to the program synthesis community, as they do not only affect bootstrapping algorithms. \\

In conclusion, bootstrapping has shown to be a powerful method to improve program synthesis performance over regular fine-tuning. There seems to be substantial promise for bootstrapping with repairing to gain additional performance over regular bootstrapping. These findings could help smaller models become better coders and enjoy a more widespread audience.

\bibliographystyle{acl_natbib}
\bibliography{references}

\appendix
\chapter{Problems with APPS}
\section{Faulty example test case}
\label{appendix:faulty_example_test_case}
\begin{lstlisting}
Question:
Given a string containing only three types of characters: '(', ')' and '*', write a function to check whether this string is valid. We define the validity of a string by these rules:

Any left parenthesis '(' must have a corresponding right parenthesis ')'.
Any right parenthesis ')' must have a corresponding left parenthesis '('.
Left parenthesis '(' must go before the corresponding right parenthesis ')'.
'*' could be treated as a single right parenthesis ')' or a single left parenthesis '(' or an empty string.
An empty string is also valid.



Example 1:

Input: "()"
Output: True



Example 2:

Input: "(*)"
Output: True



Example 3:

Input: "(*))"
Output: True



Note:

The string size will be in the range [1, 100].


class Solution:
    def checkValidString(self, s: str) -> bool:
        

Use Call-Based Format

[ANSWER]
class Solution:
	def checkValidString(self, s: str) -> bool:
		if not s:
			return True
		if s[0] == '(':
			if s[-1] == ')':
				return True
			else:
				return False
		if s[-1] == '(':
			if s[0] == ')':
				return True
			else:
				return False
		if s[0] == '*':
			if s[-1] == ')':
				return True
			else:
				return False
		if s[-1] == '*':
			if s[0] == '(':
				return True
			else:
				return False
		return True
[DONE]

Feedback: With the above function, we get the following output: [[]]. The test is "{'inputs': [' "()"\n'], 'outputs': [[{'true'}]]}". So the code does not pass the test. Please fix it.

[ANSWER]
class Solution:
	def checkValidString(self, s: str) -> bool:
		if not s:
			return True
		if s[0] == '(':
			if s[-1] == ')':
				return True
			else:
				return False
		if s[-1] == '(':
			if s[0] == ')':
				return True
			else:
				return False
		if s[0] == '*':
			if s[-1] == ')':
				return True
			else:
				return False
		if s[-1] == '*':
			if s[0] == ')':
				return True
			else:
				return False
		return True
[DONE]
\end{lstlisting}

\newpage

\section{Subset of hidden and example test cases}
This sub-selection is taken from the APPS training set, for examples that were not deemed an exact match. For the print-out we only consider tasks that have less than 3 hidden test cases. \\

It shows that many of the extracted example tests and the hidden unit tests are semantically equivalent:

\label{appendix:subset_hidden_example_test_cases} 
\begin{lstlisting}
Question: 2531
Hidden test: {'fn_name': 'findUnsortedSubarray', 'inputs': [[[2, 6, 4, 8, 10, 9, 15]]], 'outputs': [5]}
Extracted example test: {'inputs': [' [2, 6, 4, 8, 10, 9, 15]\n'], 'outputs': [' 5\nExplanation: You need to sort [6, 4, 8, 10, 9] in ascending order to make the whole array sorted in ascending order.\n']}

Question: 2532
Hidden test: {'fn_name': 'thousandSeparator', 'inputs': [[987]], 'outputs': ['987']}
Extracted example test: {'inputs': [' n = 987\n'], 'outputs': [' "987"\n']}

Question: 2534
Hidden test: {'fn_name': 'maxScore', 'inputs': [['"011101"']], 'outputs': [5]}
Extracted example test: {'inputs': [' s = "011101"\n'], 'outputs': [' 5 \nExplanation: \nAll possible ways of splitting s into two non-empty substrings are:\nleft = "0" and right = "11101", score = 1 + 4 = 5 \nleft = "01" and right = "1101", score = 1 + 3 = 4 \nleft = "011" and right = "101", score = 1 + 2 = 3 \nleft = "0111" and right = "01", score = 1 + 1 = 2 \nleft = "01110" and right = "1", score = 2 + 1 = 3\n']}

Question: 2535
Hidden test: {'fn_name': 'validPalindrome', 'inputs': [['"aba"']], 'outputs': [True]}
Extracted example test: {'inputs': [' "aba"\n'], 'outputs': [' true\n']}

Question: 2536
Hidden test: {'fn_name': 'findLucky', 'inputs': [[[2, 2, 3, 4]]], 'outputs': [2]}
Extracted example test: {'inputs': [' arr = [2,2,3,4]\n'], 'outputs': [' 2\nExplanation: The only lucky number in the array is 2 because frequency[2] == 2.\n']}

Question: 2537
Hidden test: {'fn_name': 'distanceBetweenBusStops', 'inputs': [[[1, 2, 3, 4], 0, 1]], 'outputs': [1]}
Extracted example test: {'inputs': [' distance = [1,2,3,4], start = 0, destination = 1\n'], 'outputs': [' 1\nExplanation: Distance between 0 and 1 is 1 or 9, minimum is 1.\n']}

Question: 2538
Hidden test: {'fn_name': 'countLargestGroup', 'inputs': [[13]], 'outputs': [4]}
Extracted example test: {'inputs': [' n = 13\n'], 'outputs': [' 4\nExplanation: There are 9 groups in total, they are grouped according sum of its digits of numbers from 1 to 13:\n[1,10], [2,11], [3,12], [4,13], [5], [6], [7], [8], [9]. There are 4 groups with largest size.\n']}

Question: 2539
Hidden test: {'fn_name': 'missingNumber', 'inputs': [[[3, 0, 1]]], 'outputs': [2]}
Extracted example test: {'inputs': [' [3,0,1]\n'], 'outputs': [' 2\n']}

Question: 2540
Hidden test: {'fn_name': 'largestPerimeter', 'inputs': [[[1, 2, 2]]], 'outputs': [5]}
Extracted example test: {'inputs': [' [2,1,2]\n'], 'outputs': [' 5\n']}

Question: 2542
Hidden test: {'fn_name': 'isMonotonic', 'inputs': [[[1, 2, 2, 3]]], 'outputs': [True]}
Extracted example test: {'inputs': [' [1,2,2,3]\n'], 'outputs': [' true\n']}

Question: 2543
Hidden test: {'fn_name': 'reverseOnlyLetters', 'inputs': [['"ab-cd"']], 'outputs': ['"dc-ba"']}
Extracted example test: {'inputs': [' "ab-cd"\n'], 'outputs': [' "dc-ba"\n']}

Question: 2544
Hidden test: {'fn_name': 'projectionArea', 'inputs': [[[[2], [], []]]], 'outputs': [5]}
Extracted example test: {'inputs': [' [[2]]\n'], 'outputs': [' 5\n']}   
\end{lstlisting}
\chapter{Few-shot Prompts}
\label{appendix:few_shot_examples}
\section{Regular Fine-tuning and Plain Bootstrapping Baseline Prompts}
\label{bootstrapping_baseline_prompts}
\begin{lstlisting}
### Task Start ###
Write a function to find the number of ways to fill it with 2 x 1 dominoes for the given 3 x n board.

Your code should pass these tests:
assert count_ways(2) == 3

[ANSWER]
def count_ways(n):
  A = [0] * (n + 1)
  B = [0] * (n + 1)
  A[0] = 1
  A[1] = 0
  B[0] = 0
  B[1] = 1
  for i in range(2, n+1):
    A[i] = A[i - 2] + 2 * B[i - 1]
    B[i] = A[i - 1] + B[i - 2]
  return A[n]
[DONE]

### Task End ###

### Task Start ###
Write a python function to check whether the two numbers differ at one bit position only or not.

Your code should pass these tests:
assert differ_At_One_Bit_Pos(15,8) == False

[ANSWER]
def differ_At_One_Bit_Pos(a,b):
  x = a ^ b
  return x and (not(x & (x - 1)))
[DONE]

### Task End ###

### Task Start ###
\end{lstlisting}

\section{Bootstrapping Simple Feedback Prompts}
\label{bootstrapping_simple_prompts}
\begin{lstlisting}
### Task Start ###
Write a function to find the number of ways to fill it with 2 x 1 dominoes for the given 3 x n board.

Your code should pass these tests:
assert count_ways(2) == 3

[ANSWER]
def count_ways(n):
    if n == 0:
        return 1
    if n == 1:
        return 1
    if n == 2:
        return 2
    return count_ways(n-1) + count_ways(n-2)
[DONE]

Feedback: The code above is wrong. Please fix it.

[ANSWER]
def count_ways(n):
  A = [0] * (n + 1)
  B = [0] * (n + 1)
  A[0] = 1
  A[1] = 0
  B[0] = 0
  B[1] = 1
  for i in range(2, n+1):
    A[i] = A[i - 2] + 2 * B[i - 1]
    B[i] = A[i - 1] + B[i - 2]
  return A[n]
[DONE]

Feedback: The code above is correct.

### Task End ###

### Task Start ###
Write a python function to check whether the two numbers differ at one bit position only or not.

Your code should pass these tests:
assert differ_At_One_Bit_Pos(15,8) == False

[ANSWER]
def differ_At_One_Bit_Pos(lhs,rhs):
    if (lhs - rhs) == 0 or (lhs - rhs) == 1:
        return True
    return False
[DONE]

Feedback: The code above is wrong. Please fix it.

[ANSWER]
def differ_At_One_Bit_Pos(a,b):
  x = a ^ b
  return x and (not(x & (x - 1)))
[DONE]

Feedback: The code above is correct.

### Task End ###

### Task Start ###
\end{lstlisting}

\section{Bootstrapping Interpreter Feedback Prompts}
\label{bootstrapping_interpreter_prompts}
\begin{lstlisting}
### Task Start ###
Write a function to find the number of ways to fill it with 2 x 1 dominoes for the given 3 x n board.

Your code should pass these tests:
assert count_ways(2) == 3

[ANSWER]
def count_ways(n):
    if n == 0:
        return 1
    if n == 1:
        return 1
    if n == 2:
        return 2
    return count_ways(n-1) + count_ways(n-2)
[DONE]

Feedback: With the above function, count_ways(2) == 2. The assertion is "assert count_ways(2) == 3". So the code does not pass the assertion. The code above is wrong. Please fix it.

[ANSWER]
def count_ways(n):
  A = [0] * (n + 1)
  B = [0] * (n + 1)
  A[0] = 1
  A[1] = 0
  B[0] = 0
  B[1] = 1
  for i in range(2, n+1):
    A[i] = A[i - 2] + 2 * B[i - 1]
    B[i] = A[i - 1] + B[i - 2]
  return A[n]
[DONE]

Feedback: With the above function, count_ways(2) = 3. The assertion is "assert count_ways(2) == 3". So the code passes the assertion. The code above is correct.

### Task End ###

### Task Start ###
Write a python function to check whether the two numbers differ at one bit position only or not.

Your code should pass these tests:
assert differ_At_One_Bit_Pos(15,8) == False

[ANSWER]
def differ_At_One_Bit_Pos(lhs,rhs):
    if (lhs - rhs) == 0 or (lhs - rhs) == 1:
        return True
    return False
[DONE]

Feedback: With the above function, differ_At_One_Bit_Pos(15,8) == False. The assertion is "assert differ_At_One_Bit_Pos(15,8) == False". So the code passes the assertion. However, the code above is wrong. Please fix it.

[ANSWER]
def differ_At_One_Bit_Pos(a,b):
  x = a ^ b
  return x and (not(x & (x - 1)))
[DONE]

Feedback: With the above function, differ_At_One_Bit_Pos(15,8) == False. The assertion is "assert differ_At_One_Bit_Pos(15,8) == False". So the code passes the assertion. The code above is correct.

### Task End ###

### Task Start ###
\end{lstlisting}
\chapter{Versions}
\label{appendix:versions}
Because Python versions can affect the way errors are displayed, the version of Python and some relevant libraries will be listed below. \\

We used Python 3.10.8 to train all models and run all our experiments.

\begin{table}[htbp]
\centering
\catcode`,=\active
\def,{\char`,\allowbreak}
\renewcommand\arraystretch{1.2}
\begin{tabular}{p{7cm}<{\raggedright} p{3.5cm} p{2cm}<{\raggedright} }
  \toprule
    Library           & \textbf{Version}       \\ 
    \midrule
    Torch                 &   2.0.1+cu117 \\
    Numpy                  &  1.24.2 \\
    Transformers          &   4.29.2 \\
    Tokenizers             &  0.13.2 \\
  \bottomrule
\end{tabular} 
\caption{Versions of important libraries used.}
\end{table}

For all version information, please refer to the Conda \texttt{.yml} file on the GitHub repository when it is eventually released.
\chapter{Chosen Models}
\label{appendix:chosen_models}
Here we will list which models were chosen (after how many fine-tuning steps) per training objective, based on the validation performance. Keep in mind that the fine-tuning numbers are 0-indexed. Hence, the first fine-tuning step results in model 0.

\section{MBPP}
\begin{table}[htbp]
\centering
\catcode`,=\active
\def,{\char`,\allowbreak}
\renewcommand\arraystretch{1.2}
\begin{tabular}{p{7cm}<{\raggedright} p{5cm} p{2cm}<{\raggedright} }
  \toprule
    Training objective           & \textbf{Fine-tuning step of best performing model}       \\ 
    \midrule
      Plain/regular bootstrapping & 4, 0, 1 \\
      Simple feedback bootstrapping & 3, 0, 8 \\
      Full feedback bootstrapping & 4, 7, 2 \\
  \bottomrule
\end{tabular} 
\caption{Fine-tuning step of the best performing MBPP models based on the validation performance.}
\end{table}

\section{APPS}
\begin{table}[htbp]
\centering
\catcode`,=\active
\def,{\char`,\allowbreak}
\renewcommand\arraystretch{1.2}
\begin{tabular}{p{7cm}<{\raggedright} p{5cm} p{2cm}<{\raggedright} }
  \toprule
    Training objective           & \textbf{Fine-tuning step of best performing model}       \\ 
    \midrule
      Plain/regular bootstrapping & 0, 5 \\
      Full feedback bootstrapping & 8, 2 \\
  \bottomrule
\end{tabular} 
\caption{Fine-tuning step of the best performing APPS models based on the validation performance.}
\end{table}
\chapter{Generated Code}
The validation results of the best models that have been used will be uploaded to the GitHub repository. Here a subset is displayed of some of the models on MBPP to get an idea of the model output.

\label{appendix:generated_code}
\section{MBPP}
\subsection{Regular bootstrapping model}
\subsubsection{Subset of produced code}
\begin{lstlisting}
Task 0, id=511:
### Task Start ###
Write a python function to find minimum sum of factors of a given number.
        
Your code should pass these tests:
assert find_Min_Sum(12) == 7

[ANSWER]
def find_Min_Sum(n) :
  if n == 0 :
    return 0
  if n == 1 :
    return 1
  if n % 2 == 0 :
    return find_Min_Sum(n / 2)
  else :
    return find_Min_Sum(n / 2) + find_Min_Sum(n % 2)
[DONE]
----------------------------------------------------------------------------------------------------

Task 1, id=512:
### Task Start ###
Write a function to count the element frequency in the mixed nested tuple.
        
Your code should pass these tests:
assert count_element_freq((5, 6, (5, 6), 7, (8, 9), 9) ) == {5: 2, 6: 2, 7: 1, 8: 1, 9: 2}

[ANSWER]
def count_element_freq(t):
  return dict(count_element_freq(t))
[DONE]
----------------------------------------------------------------------------------------------------

Task 2, id=513:
### Task Start ###
Write a function to convert tuple into list by adding the given string after every element.
        
Your code should pass these tests:
assert add_str((5, 6, 7, 4, 9) , "FDF") == [5, 'FDF', 6, 'FDF', 7, 'FDF', 4, 'FDF', 9, 'FDF']

[ANSWER]
def add_str(tup, str):
  return [str + x for x in tup]
[DONE]
----------------------------------------------------------------------------------------------------

Task 3, id=514:
### Task Start ###
Write a function to find the summation of tuple elements in the given tuple list.
        
Your code should pass these tests:
assert sum_elements((7, 8, 9, 1, 10, 7)) == 42

[ANSWER]
def sum_elements(tup):
  return sum(tup)
[DONE]
----------------------------------------------------------------------------------------------------

Task 4, id=515:
### Task Start ###
Write a function to check if there is a subset with sum divisible by m.
        
Your code should pass these tests:
assert modular_sum([3, 1, 7, 5], 4, 6) == True

[ANSWER]
def modular_sum(nums, m, n):
  if (m % n == 0):
    return True
  else:
    return False
[DONE]
----------------------------------------------------------------------------------------------------
\end{lstlisting}

\subsubsection{Subset of correctly produced code}
\begin{lstlisting}
Correct codes
Task 0, id=514:
Task: ### Task Start ###
Write a function to find the summation of tuple elements in the given tuple list.
        
Your code should pass these tests:
assert sum_elements((7, 8, 9, 1, 10, 7)) == 42

[ANSWER]
def sum_elements(tup):
  return sum(tup)
[DONE]
----------------------------------------------------------------------------------------------------

Correct codes
Task 1, id=516:
Task: ### Task Start ###
Write a function to sort a list of elements using radix sort.
        
Your code should pass these tests:
assert radix_sort([15, 79, 25, 68, 37]) == [15, 25, 37, 68, 79]

[ANSWER]
def radix_sort(list1):
  list1.sort()
  return list1
[DONE]
----------------------------------------------------------------------------------------------------

Correct codes
Task 2, id=517:
Task: ### Task Start ###
Write a python function to find the largest postive number from the given list.
        
Your code should pass these tests:
assert largest_pos([1,2,3,4,-1]) == 4

[ANSWER]
def largest_pos(nums):
  return max(nums)
[DONE]
----------------------------------------------------------------------------------------------------

Correct codes
Task 3, id=532:
Task: ### Task Start ###
Write a function to check if the two given strings are permutations of each other.
        
Your code should pass these tests:
assert check_permutation("abc", "cba") == True

[ANSWER]
def check_permutation(a, b):
  return sorted(a) == sorted(b)
[DONE]
----------------------------------------------------------------------------------------------------

Correct codes
Task 4, id=541:
Task: ### Task Start ###
Write a function to find if the given number is abundant or not.
        
Your code should pass these tests:
assert check_abundant(12) == True

[ANSWER]
def check_abundant(n) :
  if n % 2 == 0 or n == 2 :
    return True
  for i in range(3, int(n**0.5)+1, 2) :
    if n % i == 0 or n == i :
      return False
  return True
[DONE]
----------------------------------------------------------------------------------------------------
\end{lstlisting}

\subsection{Full feedback repair model}
\subsubsection{Subset of produced code}
\begin{lstlisting}
Task 0, id=511:
### Task Start ###
Write a python function to find minimum sum of factors of a given number.
        
Your code should pass these tests:
assert find_Min_Sum(12) == 7

[ANSWER]
def find_Min_Sum(n):
	if n == 1:
		return 1
	else:
		return find_Min_Sum(n / 2) + find_Min_Sum(n / 3)
[DONE]
----------------------------------------------------------------------------------------------------



Task 1, id=512:
### Task Start ###
Write a function to count the element frequency in the mixed nested tuple.
        
Your code should pass these tests:
assert count_element_freq((5, 6, (5, 6), 7, (8, 9), 9) ) == {5: 2, 6: 2, 7: 1, 8: 1, 9: 2}

[ANSWER]
def count_element_freq(tup):
	res = {}
	for i in range(len(tup)):
		res[tup[i]] = tup[i]
	return (res)
[DONE]
----------------------------------------------------------------------------------------------------



Task 2, id=513:
### Task Start ###
Write a function to convert tuple into list by adding the given string after every element.
        
Your code should pass these tests:
assert add_str((5, 6, 7, 4, 9) , "FDF") == [5, 'FDF', 6, 'FDF', 7, 'FDF', 4, 'FDF', 9, 'FDF']

[ANSWER]
def add_str(tup, s):
	res = [str(x) + s for x in tup]
	return (res)
[DONE]
----------------------------------------------------------------------------------------------------



Task 3, id=514:
### Task Start ###
Write a function to find the summation of tuple elements in the given tuple list.
        
Your code should pass these tests:
assert sum_elements((7, 8, 9, 1, 10, 7)) == 42

[ANSWER]
def sum_elements(tup):
	return sum(tup)
[DONE]
----------------------------------------------------------------------------------------------------



Task 4, id=515:
### Task Start ###
Write a function to check if there is a subset with sum divisible by m.
        
Your code should pass these tests:
assert modular_sum([3, 1, 7, 5], 4, 6) == True

[ANSWER]
def modular_sum(arr, m, n):
	for i in range(m):
		for j in range(n):
			if arr[i] + arr[j] % m == 0:
				return True
	return False
[DONE]
----------------------------------------------------------------------------------------------------
\end{lstlisting}

\subsubsection{Subset of correctly produced code}
\begin{lstlisting}
Correct codes
Task 0, id=514:
Task: ### Task Start ###
Write a function to find the summation of tuple elements in the given tuple list.
        
Your code should pass these tests:
assert sum_elements((7, 8, 9, 1, 10, 7)) == 42

[ANSWER]
def sum_elements(tup):
	return sum(tup)
[DONE]
----------------------------------------------------------------------------------------------------


Correct codes
Task 1, id=516:
Task: ### Task Start ###
Write a function to sort a list of elements using radix sort.
        
Your code should pass these tests:
assert radix_sort([15, 79, 25, 68, 37]) == [15, 25, 37, 68, 79]

[ANSWER]
def radix_sort(arr):
	arr.sort()
	return arr
[DONE]
----------------------------------------------------------------------------------------------------

Correct codes
Task 2, id=517:
Task: ### Task Start ###
Write a python function to find the largest postive number from the given list.
        
Your code should pass these tests:
assert largest_pos([1,2,3,4,-1]) == 4

[ANSWER]
def largest_pos(nums):
	max_pos = max(nums)
	for i in range(1, len(nums)):
		if nums[i] > max_pos:
			max_pos = nums[i]
	return max_pos
[DONE]
----------------------------------------------------------------------------------------------------

Correct codes
Task 3, id=533:
Task: ### Task Start ###
Write a function to remove particular data type elements from the given tuple.
        
Your code should pass these tests:
assert remove_datatype((4, 5, 4, 7.7, 1.2), int) == [7.7, 1.2]

[ANSWER]
def remove_datatype(test_tup, data_type):
	res = [x for x in test_tup if type(x)!= data_type]
	return (res)
[DONE]
----------------------------------------------------------------------------------------------------

Correct codes
Task 4, id=539:
Task: ### Task Start ###
Write a function to create a list containing the power of said number in bases raised to the corresponding number in the index using map function.
        
Your code should pass these tests:
assert basesnum_coresspondingnum([10, 20, 30, 40, 50, 60, 70, 80, 90, 100],[1, 2, 3, 4, 5, 6, 7, 8, 9, 10])==[10, 400, 27000, 2560000, 312500000, 46656000000, 8235430000000, 1677721600000000, 387420489000000000, 100000000000000000000]

[ANSWER]
def basesnum_coresspondingnum(nums,nums2):
	res = [pow(x,y) for x,y in zip(nums,nums2)]
	return (res)
[DONE]
----------------------------------------------------------------------------------------------------
\end{lstlisting}

\subsubsection{Subset of incorrect repair attempts}
\begin{lstlisting}
Incorrect codes
Task 0, id=511:
### Task Start ###
Write a python function to find minimum sum of factors of a given number.
        
Your code should pass these tests:
assert find_Min_Sum(12) == 7

[ANSWER]
def find_Min_Sum(n):
	if n == 1:
		return 1
	else:
		return find_Min_Sum(n / 2) + find_Min_Sum(n / 3)
[DONE]

Feedback: With the above function, find_Min_Sum(12) returns the following error:
"""
RecursionError: maximum recursion depth exceeded in comparison
"""
So the code does not pass the assertion. Please fix it.

[ANSWER]
def find_Min_Sum(n):
	if n == 1:
		return 1
	else:
		return find_Min_Sum(n - 1) + find_Min_Sum(n - 2)
[DONE]
----------------------------------------------------------------------------------------------------



Task 1, id=512:
### Task Start ###
Write a function to count the element frequency in the mixed nested tuple.
        
Your code should pass these tests:
assert count_element_freq((5, 6, (5, 6), 7, (8, 9), 9) ) == {5: 2, 6: 2, 7: 1, 8: 1, 9: 2}

[ANSWER]
def count_element_freq(tup):
	res = {}
	for i in range(len(tup)):
		res[tup[i]] = tup[i]
	return (res)
[DONE]

Feedback: With the above function, count_element_freq((5, 6, (5, 6), 7, (8, 9), 9) ) == {5: 5, 6: 6, (5, 6): (5, 6), 7: 7, (8, 9): (8, 9), 9: 9}. The assertion is "assert count_element_freq((5, 6, (5, 6), 7, (8, 9), 9) ) == {5: 2, 6: 2, 7: 1, 8: 1, 9: 2}". So the code does not pass the assertion. Please fix it.

[ANSWER]
def count_element_freq(tup):
	res = {}
	for i in range(len(tup)):
		res[tup[i]] = res.get(tup[i], 0) + 1
	return (res)
[DONE]
----------------------------------------------------------------------------------------------------



Task 2, id=513:
### Task Start ###
Write a function to convert tuple into list by adding the given string after every element.
        
Your code should pass these tests:
assert add_str((5, 6, 7, 4, 9) , "FDF") == [5, 'FDF', 6, 'FDF', 7, 'FDF', 4, 'FDF', 9, 'FDF']

[ANSWER]
def add_str(tup, s):
	res = [str(x) + s for x in tup]
	return (res)
[DONE]

Feedback: With the above function, add_str((5, 6, 7, 4, 9) , "FDF") == ['5FDF', '6FDF', '7FDF', '4FDF', '9FDF']. The assertion is "assert add_str((5, 6, 7, 4, 9) , "FDF") == [5, 'FDF', 6, 'FDF', 7, 'FDF', 4, 'FDF', 9, 'FDF']". So the code does not pass the assertion. Please fix it.

[ANSWER]
def add_str(tup, s):
	res = [str(x) + s for x in tup]
	return (res)
[DONE]
----------------------------------------------------------------------------------------------------



Task 3, id=515:
### Task Start ###
Write a function to check if there is a subset with sum divisible by m.
        
Your code should pass these tests:
assert modular_sum([3, 1, 7, 5], 4, 6) == True

[ANSWER]
def modular_sum(arr, m, n):
	for i in range(m):
		for j in range(n):
			if arr[i] + arr[j] % m == 0:
				return True
	return False
[DONE]

Feedback: With the above function, modular_sum([3, 1, 7, 5], 4, 6) returns the following error:
"""
IndexError: list index out of range
"""
So the code does not pass the assertion. Please fix it.

[ANSWER]
def modular_sum(arr, m, n):
	for i in range(m):
		for j in range(n):
			if (arr[i] + arr[j] % m == 0):
				return True
	return False
[DONE]
----------------------------------------------------------------------------------------------------



Task 4, id=518:
### Task Start ###
Write a function to find the square root of a perfect number.
        
Your code should pass these tests:
assert sqrt_root(4)==2

[ANSWER]
def sqrt_root(n):
	if n == 0:
		return 0
	if n == 1:
		return 1
	if n % 2 == 0:
		return sqrt_root(n / 2)
	else:
		return sqrt_root(n / 2) + sqrt_root(n % 2)
[DONE]

Feedback: With the above function, sqrt_root(400) returns the following error:
"""
RecursionError: maximum recursion depth exceeded in comparison
"""
So the code does not pass the assertion. Please fix it.

[ANSWER]
def sqrt_root(n):
	if n == 0:
		return 0
	if n == 1:
		return 1
	if n % 2 == 0:
		return sqrt_root(n / 2)
	else:
		return sqrt_root(n / 2) + sqrt_root(n % 2)
[DONE]
\end{lstlisting}

\subsubsection{Correctly repaired code}
\begin{lstlisting}
Correct repaired codes
Task 0, id=563:
### Task Start ###
Write a function to extract values between quotation marks of a string.
        
Your code should pass these tests:
assert extract_values('"Python", "PHP", "Java"')==['Python', 'PHP', 'Java']

[ANSWER]
def extract_values(s):
	res = s.split('"')
	res = [x.strip() for x in res]
	return (res)
[DONE]

Feedback: With the above function, extract_values('"Python", "PHP", "Java"')==['', 'Python', ',', 'PHP', ',', 'Java', '']. The assertion is "assert extract_values('"Python", "PHP", "Java"')==['Python', 'PHP', 'Java']". So the code does not pass the assertion. Please fix it.

[ANSWER]
import re
def extract_values(s):
	res = re.findall('\"(.*?)\"', s)
	res = [x.strip() for x in res]
	return (res)
[DONE]
----------------------------------------------------------------------------------------------------



Task 1, id=591:
### Task Start ###
Write a python function to interchange the first and last elements in a list.
        
Your code should pass these tests:
assert swap_List([12, 35, 9, 56, 24]) == [24, 35, 9, 56, 12]

[ANSWER]
def swap_List(list1):
	list1.insert(0, list1.pop())
	list1.append(list1.pop())
	return list1
[DONE]

Feedback: With the above function, swap_List([12, 35, 9, 56, 24]) == [24, 12, 35, 9, 56]. The assertion is "assert swap_List([12, 35, 9, 56, 24]) == [24, 35, 9, 56, 12]". So the code does not pass the assertion. Please fix it.

[ANSWER]
def swap_List(list1):
	list1[0], list1[-1] = list1[-1], list1[0]
	return list1
[DONE]
----------------------------------------------------------------------------------------------------



Task 2, id=599:
### Task Start ###
Write a function to find sum and average of first n natural numbers.
        
Your code should pass these tests:
assert sum_average(10)==(55, 5.5)

[ANSWER]
def sum_average(n):
	sum = 0
	avg = 0
	for i in range(n):
		sum += i
		avg += i
	avg = avg / n
	return (sum, avg)
[DONE]

Feedback: With the above function, sum_average(10)==(45, 4.5). The assertion is "assert sum_average(10)==(55, 5.5)". So the code does not pass the assertion. Please fix it.

[ANSWER]
def sum_average(n):
	sum = 0
	avg = 0
	for i in range(1, n + 1):
		sum += i
		avg += i
	avg = avg / n
	return (sum, avg)
[DONE]
----------------------------------------------------------------------------------------------------
\end{lstlisting}

\section{APPS}
The APPS results will be released on the GitHub repository as the prompts are extremely large and would fill up the appendix.

\end{document}